lsl

\documentclass[smallextended]{svjour3}       % onecolumn (second format)
\smartqed  % flush right qed marks, e.g. at end of proof
\usepackage{graphicx}
\usepackage{amssymb}
\usepackage{amsmath}
\usepackage{bm}
\usepackage{wasysym}
\usepackage{lscape}
\usepackage{natbib}
\begin{document}

\title{Influence of external forcings on abrupt millennial-scale climate changes: a statistical modelling study
%A statistical modelling study of the abrupt millennial-scale climate changes focusing on the influence of external forcings%\thanks{Grants or other notes
}

\titlerunning{Influence of external forcings on abrupt millennial-scale climate changes}        % if too long for running head

\author{Takahito Mitsui        \and
        Michel Crucifix %etc.
}


\institute{T. Mitsui \at
              Universit\'e catholique de Louvain, Earth and Life Institute, Georges Lema\^\i tre Centre for Earth and Climate Research, BE-1348 Louvain-la-Neuve, Belgium \\
              Tel.: +32-10-473236\\
%             Fax: +123-45-678910\\
              \email{takahito321@gmail.com}           %  \\
           \and
           M. Crucifix \at
              Universit\'e catholique de Louvain, Earth and Life Institute, Georges Lema\^\i tre Centre for Earth and Climate Research, BE-1348 Louvain-la-Neuve, Belgium, \\
Belgian National Fund of Scientific Research, Rue d'Egmont, 5
BE-1000 Brussels, Belgium
}

\date{Received: 21 October 2015 / Accepted: 10 June 2016 in Climate Dynamics}

\maketitle
%\begin{linenumbers}

\begin{abstract}
The last glacial period was punctuated by a series of abrupt climate shifts, the so-called Dansgaard-Oeschger (DO) events. The frequency of DO events varied in time, supposedly because of changes in background climate conditions. Here, the influence of external forcings on DO events is investigated with statistical modelling. We assume two types of simple stochastic dynamical systems models (double-well potential-type and oscillator-type), forced by the northern hemisphere summer insolation change and/or the global ice volume change. The model parameters are estimated by using the maximum likelihood method with the NGRIP Ca$^{2+}$ record. The stochastic oscillator model with at least the ice volume forcing reproduces well the sample autocorrelation function of the record and the frequency changes of warming transitions in the last glacial period across MISs 2, 3, and 4. The model performance is improved with the additional insolation forcing. The BIC scores also suggest that the ice volume forcing is relatively more important than  the insolation forcing, though the strength of evidence depends on the model assumption. Finally, we simulate the average number of warming transitions in the past four glacial periods, assuming the model can be extended beyond the last glacial, and compare the result with an Iberian margin sea-surface temperature (SST) record (Martrat {\it et al.}, Science, vol.~317, p.~502, 2007). The simulation result supports the previous observation that abrupt millennial-scale climate changes in the penultimate glacial (MIS~6) are less frequent than in the last glacial (MISs~2--4). On the other hand, it suggests that the number of abrupt millennial-scale climate changes in older glacial periods (MISs 6, 8, and 10) might be larger than inferred from the SST record. 
\keywords{Dansgaard-Oeschger events \and abrupt millennial-scale climate changes \and statistical modelling \and orbital insolation forcing \and global ice volume change}
\PACS{92.70.Aa \and 92.70.Gt \and 05.45.Tp}
\end{abstract}

\section{Introduction}
\begin{figure}
\begin{center}
  \includegraphics[angle=270,width=0.95\linewidth]{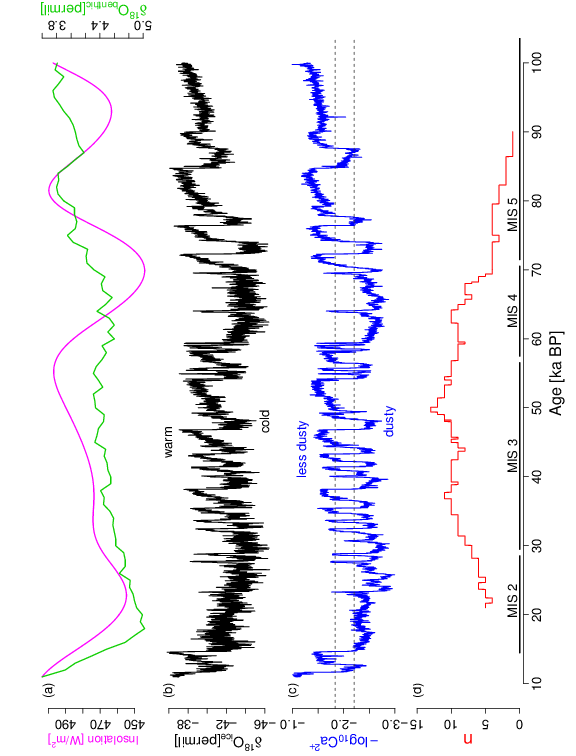}
  \caption{Comparison between NGRIP ice core records and background climate conditions. (a) The mean monthly insolation from 21 June to 20 July at 65$^{\circ}$N (magenta) \citep{laskar2004long} and the benthic oxygen isotope ratio $\delta ^{18}$O$_{\text{benthic}}$ as a proxy for the global ice volume (green) \citep{lisiecki2005pliocene}. (b) The  20-year average NGRIP $\delta ^{18}$O$_{\text{ice}}$ \citep{rasmussen2014stratigraphic}. (c) The 20-year average NGRIP Ca$^{2+}$ concentration (ppb) (blue) \citep{rasmussen2014stratigraphic}. The dashed lines are the upper and lower thresholds used to define a warming transition. (d) The number of warming transitions $n(t)$ for each moving time window [$t-10$~ka, $t+10$~ka]. The marine isotope stages (MISs) are based on \cite{lisiecki2005pliocene}.} \label{fig:intro}
\end{center}
\end{figure}

During the last glacial period, the North Atlantic region experienced a series of abrupt climate shifts between cold (stadial) and relatively warm (interstadial) phases, the so-called Dansgaard-Oeschger (DO) events \citep{dansgaard1993evidence}. These are clearly reflected in changes in the oxygen isotope ratio $\delta ^{18}$O$_\text{ice}$ (a proxy for air temperature) of Greenland ice cores (see Fig.~\ref{fig:intro}(b)). 
% \cite{rasmussen2014stratigraphic} document 25 major warming events and several sub-events during the last glacial cycle. 
Typically, abrupt warmings occurred within a few decades, and they were followed by a gradual cooling before a rapid return to a cold state. The amplitude of the abrupt warmings ranges from $8^{\circ}$C to $16^{\circ}$C (\cite{wolff2010millennial} and references therein).

DO events are commonly associated with changes in deep-ocean activity and sea-ice cover in the North Atlantic \citep{gildor2003sea,denton2005role,li2005abrupt}. They are also associated with changes in the large scale thermohaline circulation (THC) \citep{broecker1985does,rahmstorf2002ocean}, though such circulation changes do not seem as dramatic as those that occurred during Heinrich events (\cite{elliot2002changes,clement2008mechanisms} and references threrein). What causes the onset, demise, and recurrence of DO events is still not so clear \citep{clement2008mechanisms}. A number of modelling studies show that the convective activity and the broader THC depend nonlinearly on the freshwater balance of the North Atlantic \citep{manabe1988two,ganopolski2001rapid}. In turn, such circulation changes may impact the mass balance of the surrounding ice sheets and their freshwater supply onto the ocean. Such interplay may explain complex dynamics of DO events \citep{kageyama2005dansgaard}. Alternatively, it has been suggested that self-sustained oscillations are possible as a result of advective and convective dynamics in the ocean without any change in freshwater input (for example, \cite{colin2007simple}).

Greenland ice cores contain various continental dusts transported from mainly East Asian deserts \citep{biscaye1997asian}. There is a strong correlation between $\delta ^{18}$O$_\text{ice}$ (Fig.~\ref{fig:intro}(b)) and dust concentrations (approximated by [Ca$^{2+}$], Fig.~\ref{fig:intro}(c)) in the ice cores. This suggests that North Atlantic climate changes are tightly linked with changes in northern hemisphere atmospheric circulation \citep{mayewski1997major} or dust storm activity in East Asia \citep{ruth2007ice}. A recent simulation shows that the increase in the meridional temperature gradient in the North Atlantic leads to stronger westerlies (important for long-range dust transport) and strengthened winter wind speed above major Asian dust source regions (important for dust entrainment) \citep{sun2012influence}.   

The occurrence frequency of DO events varied in time as shown in Fig.~\ref{fig:intro}. The frequency is quantified by counting the number of warming transitions $n(t)$ for each 20-ka moving window [$t-10$~ka, $t+10$~ka] over the 20-year average NGRIP Ca$^{2+}$ record \citep{rasmussen2014stratigraphic}. Here, we standardize $-\log _{10}[\text{Ca}^{2+}]$ by the mean and standard deviation during 11--100~ka~BP as $y=(-\log _{10}[\text{Ca}^{2+}]+2.02)/0.464$. A warming transition is then defined as the first up-crossing of the upper threshold\footnote{The thresholds are set to $y=\pm 0.4$ to be able to count the lowest interstadial event with $y\sim 0.5$ (GI-4 in \cite{rasmussen2014stratigraphic}) and the highest stadial event with $y\sim -0.7$ (GS-22 in the same). These thresholds yield total 26 warming transitions during 11--100~ka~BP. Similar criteria are used in \citep{alley2001stochastic,ditlevsen2005climate}.} $y=0.4$ after falling below the lower threshold $y=-0.4$ (dashed lines in Fig.~\ref{fig:intro}(c)). The number of warming transitions $n(t)$ increases from MIS~5 to MIS~3 and decreases from MIS~3 to MIS~2 (Fig.~\ref{fig:intro}(d)). The purpose of this paper is to explore external forcings which induce these frequency changes. 
%The purpose of this paper is explore possible external forcings on DO events, which induce these frequency changes.
%The frequency changes of DO events may be related to slowly-changing background climate components:
%\footnote{Here, we are not mentioning about possible {\it triggers} of DO events, which may act in much faster time scales, such as weather forcing \citep{monahan2008stochastic} or solar forcing \citep{goosse2002potential,braun2008solar}.}:

{\it Ice sheet forcing}. The millennial-scale climate changes have larger amplitudes when the global ice volume is in intermediate level \citep{mcmanus19990}. Northern hemisphere ice sheets may affect the THC or the sea ice formation in the North Atlantic by their meltwater discharges \citep{schulz2002relaxation,knutti2004strong,jackson2010box}\footnote{\cite{schulz2002relaxation} assume that the North Atlantic THC is controlled by the freshwater flux anomaly (runoff from ice sheets) in proportion to the ice volume itself, referring to \cite{marshall1999modeling}. However, in other studies, the freshwater flux is related to the loss of the ice volume \citep{knutti2004strong,jackson2010box}.}, by changing wind fields \citep{wunsch2006abrupt}, or by their albedo effect. Using a comprehensive climate model, \cite{zhang2014abrupt} show that small changes of the height of the northern hemisphere ice sheets can cause rapid climate transitions like DO events via an atmosphere-ocean-sea-ice feedback in the North Atlantic. 

{\it Astronomical insolation forcing}. The seasonal and latitudinal distribution of the insolation at the top of the atmosphere varies due to the long term variations of the Earth's astronomical parameters: climatic precession, obliquity, and eccentricity \citep{berger1991insolation}. Several observational as well as modelling studies propose the influence of boreal summer insolation change (Fig.~\ref{fig:intro}(a)) on DO events \citep{adams1999sudden,martrat2004abrupt,rial2007frequency,capron2010millennial}.
%The mean monthly insolation from 21 June to 20 July at 65$^{\circ}$N changed by 59~W/m$^{2}$ in the period 11--100~ka~BP as shown in Fig.~\ref{fig:intro}(a) \citep{laskar2004long}. This change amounts to $\sim $12\% of the temporal average of the mean monthly insolation 475~W/m$^2$ during the period 11--100~ka~BP. \cite{ve1ez2001stochastic} considered the influence of such an astronomical forcing on the THC, referring to \cite{adams1999sudden}. 
%\cite{rial2007frequency} propose that the frequency of DO oscillations is modulated by the northern hemisphere summer (NHS) insolation change. 
%Their simulation results closely resemble observed $\delta ^{18}$O$_{\text{ice}}$ records.
% \cite{martrat2004abrupt} as well as \cite{capron2010millennial}\footnote{In particular, \cite{capron2010millennial} suggest a link between NHS insolation and the presence of a sub-millennial scale climatic variability, specifically a precursor-type event before a DO event.} also mention the influence of NHS insolation change, which can affect temperature, seasonality, hydrological cycle, and ice sheet growth in the high latitudes. 
On the other hand, \cite{masson2005grip,olsen2005ocean,friedrich2010mechanism} emphasize the influence of the latitudinal gradient of annual mean insolation (i.e., obliquity forcing). 

{\it CO$_2$ forcing}. The atmospheric CO$_2$ concentration varied in a range of 180--280~ppm over the last four glacial cycles \citep{petit1999climate}. 
%This amounts to the change in the radiative forcing by $5.85\ln (280/180)\sim2.4$~W/m$^2$. 
Increases of CO$_2$ by $\sim $20~ppm are observed prior to large DO warmings after Heinrich events \citep{ahn2008atmospheric}. 
%A change of CO$_2$ from 185~p.p.m. to 205~p.p.m. increases the radiative forcing by 0.55~W/m$^2$ \citep{myhre1998new}. This is a relatively small value, but 
A fully coupled model simulation by \cite{zhang2014abrupt} shows an increase of CO$_2$ by 20~ppm (corresponding to the change in the radiative forcing by $\sim$0.55~W/m$^2$) can trigger warming transitions like DO events.
%{\it Solar variability}. The variations of the total solar irradiance (TSI) over the past 400~years are estimated as less than $\sim$0.2\% of the total \citep{myhre2013co}. This is a small value, but it must be noted that changes as small as $\sim$0.4\% of the TSI have been seen to trigger abrupt events in a low-resolution 3-D model \citep{goosse2002potential}. It has been proposed that changes in TSI may induce a periodic response in the climate system. Specifically, \cite{braun2008solar} suggested that centennial solar cycles (the $\sim 87$~yr DeVriesSuess and the $\sim 210$~yr Gleissberg cycle) may be involved in a mechanism of ``ghost stochastic resonance", as an explanation to the observation that Dansgaard-Oeschger events occurred with periods multiple of 1470~yr \citep{schul

In this study, we focus on the northern hemisphere summer insolation forcing and the ice volume forcing. The ice volume contains the information of insolation change in part \citep{hays76}. However, with close inspection in Fig.~\ref{fig:intro}(a), these curves have only a weak correlation (the coefficient of determination $R^2=0.14$). The CO$_2$ forcing is not explicitly considered here, but it is implicitly included in the ice volume forcing because the atmospheric CO$_2$ concentration (in logarithmic scale) is strongly correlated to the global sea-level ($R^2=0.68$ over the past 550~ka \citep{foster2013relationship}). There is a debate on the temporal regularity of DO events. \cite{schulz20021470} proposed that DO events occurred with periods multiple of 1470~yr. The periodicity evoked the existence of external clock, for example, solar cycles \citep{braun2008solar} or tidal cycles \citep{keeling20001}.
%\cite{braun2008solar} suggested that centennial solar cycles (the $\sim 87$~yr DeVriesSuess and the $\sim 210$~yr Gleissberg cycle) may be involved in a mechanism of ``ghost stochastic resonance", as an explanation to \citep{schulz20021470}.  
However, the statistical significance of 1470-yr periodicity is questioned by \cite{ditlevsen2007climate}. He argues that the onsets of DO events are indistinguishable from a random occurrence (more precisely a Poisson process). Thus, we assume that millennial cycles are generated by stochastic dynamics without explicit 1470-yr forcing.

To infer the influence of external forcings in noisy records, we take an approach by statistical modelling based on dynamical systems, which has many paleoclimatic applications: ice ages \citep{hargreaves2002assimilation,crucifix2009use}, DO events \citep{kwasniok2009deriving,kwasniok2012stochastic,peavoy2010bayesian,kwasniok2013analysis}, and abrupt monsoon transitions \citep{thomas2015early}.
%, where model parameters are estimated by the maximum likelihood method with the state estimation by the unscented Kalman filter \citep{julier2004unscented} before the models themselves are compared using the Bayesian Information Criterion (BIC). 
%Following \cite{kwasniok2013analysis}, 
As in \cite{kwasniok2013analysis}, we assume two types of simple dynamical systems models based on two paradigms of DO dynamics, {\it bistability} and {\it oscillations}, but here these systems are forced by the northern hemisphere summer insolation change and/or the ice volume change. %The combination of two dynamical systems and two external forcings yields 8 sub-models. 
In our analysis, superiority of the oscillator model and the relative importance of the ice volume forcing are shown. Finally, we simulate the abrupt millennial-scale climate changes beyond the last glacial, assuming the model can be extended, and compare the result with the Iberian margin SST record over these periods \citep{martrat2007four}. 
%It is challenging to reconstruct abrupt millennial-scale climate changes before the Eemian interglacial because the Greenland records are severely disturbed by effects caused by the basal flow. The information earlier than the Eemian is found in records such as Iberian margin sea-surface temperatures (SSTs) records \citep{martrat2007four}, pollen records \citep{margari2010nature}, and speleothem Asian monsoon records \citep{cheng2009ice}. These records suggest that the abrupt millennial-scale climate variability is an ubiquitous feature over the past several glacial periods \citep{mcmanus19990,barker2011800}. However, it seems more difficult to identify abrupt climate changes in marine and terrestrial records than in high-resolution Greenland records. In this paper, we simulate the occurrence frequency of the abrupt climate changes in the past four glacial periods using a model calibrated with the ice core record in the last glacial cycle, and compare the result with the Iberian margin SST record over these periods \citep{martrat2007four}.

The remainder of this article is organized as follows. In Section~2, we explain the data and methods. In Section~3, we show the results of the model parameter estimation. The calibrated models are compared in Section~4. We predict the occurrence frequency of abrupt millennial-scale climate changes in the past four glacial periods in Section~5.

\section{Data and methods}
%We describe climatic data, statistical models, state and parameter estimation methods, and the model comparison method used in this study.
\subsection{Data}
%As mentioned in Introduction, $\delta ^{18}$O$_{\text{ice}}$ is commonly used as a proxy for air temperature, and 
The calcium ion concentration [Ca$^{2+}$] in ice cores is a proxy of continental dusts transported from mainly East Asian deserts, which depend on several factors \citep{fuhrer1999timescales}: (i) source area conditions (such as aridity or vegetation), (ii) mobilization of dusts by winds in source areas and uplift to transporting levels, (iii) long-range transport efficiency depending on transient times, (iv) losses en route (gravitational settling or wash-out), and (v) deposition on ice sheet. The relative contribution is a matter of debate. \cite{fischer2007glacial} propose larger contributions of both source strength and transport, while \cite{ruth2007ice} propose that the increase of dust concentrations during stadials is largely attributed to the increased dust storm activity in East Asia.

The quantity $-\log _{10}$[Ca$^{2+}$] is well correlated with $\delta ^{18}$O$_{\text{ice}}$ ($R^2=0.82$ in the case of Fig.~\ref{fig:intro}), and it is assumed to change synchronously with the oxygen isotope ratio $\delta ^{18}$O$_{\text{ice}}$ within 20-year resolution \citep{rasmussen2014stratigraphic}.
%Note that stadials and interstadials of $-\log _{10}$[Ca$^{2+}$] are more symmetric than those of $\delta ^{18}$O$_{\text{ice}}$ (Fig.~\ref{fig:intro}) \citep{barker2007antarctic}. 
There are advantages in the use of Ca$^{2+}$ signals: Ca$^{2+}$ has an excellent signal-to-noise ratio \citep{rasmussen2014stratigraphic}, and the differences between ice cores (NGRIP, GRIP, and GISP2) are small compared to $\delta ^{18}$O$_{\text{ice}}$ (see Fig.~1 in \cite{rasmussen2014stratigraphic}).

We use the 20-year average NGRIP Ca$^{2+}$ on the GICC05modelext timescale \citep{rasmussen2014stratigraphic,bigler2004hochauflosende} standardized as $y=(-\log _{10}[\text{Ca}^{2+}]+2.02)/0.464$ by the mean and standard deviation during $11$--$100$~ka~BP\footnote{The 20-year average NGRIP Ca$^{2+}$ record provided in \cite{seierstad2014consistently} has 26 missing values during $11$--$100$~ka~BP, which represent 0.6\% of the 4451 data points. We just interpolate them linearly for simplicity.}. To assess the proxy dependence of the result, we also use the 20-year average NGRIP oxygen isotope ratio $\delta ^{18}$O$_{\text{ice}}$ \citep{rasmussen2014stratigraphic} standardized in a similar manner.

In our models (next Subsection), we employ the mean monthly insolation from 21 June to 20 July at 65$^{\circ}$N \citep{laskar2004long} and the global ice volume estimated from the benthic oxygen isotope ratio $\delta ^{18}$O$_{\text{benthic}}$ (the LR04 stack record by \cite{lisiecki2005pliocene}). The former is standardized by subtracting the mean 474.93~W/m$^2$ and dividing by standard deviation 15.08~W/m$^2$ during 11--100~ka~BP. This is referred to as the insolation forcing $I(t)$. The LR04 stack record is linearly interpolated to have $20$-yr resolution and then standardized with the mean 4.34~$\permil$ and standard deviation 0.33~$\permil$. This is referred to as the ice volume forcing $V(t)$. $\delta ^{18}$O$_{\text{benthic}}$ is affected both by the global ice volume and deep water temperature, but we use $\delta ^{18}$O$_{\text{benthic}}$ as a rough approximation of the global ice volume change. %The LR04 stack record has a resolution of $1$~ka in the last $100$~ka so that we linearly interpolate the data to have $20$-yr resolution. 

\subsection{Models}
Given that the mechanism of DO events is still in debate, we assume simple abstract dynamical models: a stochastic one-dimensional (1D) potential model and a stochastic oscillator model \citep{kwasniok2013analysis}. %These models are forced by the NHS insolation change $I(t)$ and the global ice volume change $V(t)$. 
The way to include the forcings $I(t)$ and $V(t)$ in the models is not obvious. As the simplest assumption, we assume that the system is linearly forced by $I(t)$ and $V(t)$ with relative weights $\gamma _1$ and $\gamma _2$.  

\subsubsection{Stochastic 1D potential model}
The most prevalent hypothesis is that DO events are the transitions between two climate states corresponding to ``warm'' and ``cold'' modes of the THC \citep{broecker1985does,rahmstorf2002ocean}. Denote the model's ``true'' state for the standardized $-\log _{10}$[Ca$^{2+}$] by $x(t)$. The bistability hypothesis can be expressed by the following stochastic dynamical systems model:   
\begin{equation}
\begin{split}
&dx(t)=[-U'(x) + \gamma _1 I(t) -\gamma _2 V(t)]dt +\sigma dW(t),\\
&U(x)=a_1x+a_2x^2+a_3x^3+a_4x^4,\,\,\,a_4>0,
\end{split} \label{eq:1DPo}
\end{equation}
where $\sigma dW(t)$ is the system noise and $W(t)$ is the standard Wiener process \citep{gardinerstochastic}. The infinitesimal increment $dW(t)=W(t+dt)-W(t)$ obeys a normal distribution $\mathcal{N}(0,dt)$ with mean zero and variance $dt$. The 4th order potential $U(x)$ is a minimal model which allows bimodality. Equations similar to (\ref{eq:1DPo}) have been used to describe the dynamics of THC \citep{stommel1993average,cessi1994simple,timmermann2000noise,ve1ez2001stochastic,monahan2006probability}, DO events \citep{ditlevsen1999observation,kwasniok2009deriving,rial2011modeling}, or abrupt monsoon transitions \citep{thomas2015early}.

The measurement $y(t)$ obeys the following observation model:
\begin{equation}
y(t)=x(t)+\eta (t),\,\,\,\eta (t)\sim \mathcal{N}(0,\varepsilon ^2). \label{eq:obs}
\end{equation}
The observation noise $\eta (t)$ includes all the discrepancies between the model state and the measurement (i.e., instrument errors and representation errors). 
It is modeled by a white noise obeying a normal distribution $\mathcal{N} (0,\varepsilon ^2)$.
In addition to this original model with 8 parameters, we consider a submodel with 7 parameters obtained by setting $\varepsilon\equiv 0$. We refer to the original model as the 1D potential model A and the submodel as the 1D potential model B.

\subsubsection{Stochastic oscillator model}
The second paradigm of DO events is based on {\it self-sustained oscillations} \citep{broecker1990salt}, which appear in simple conceptual models \citep{birchfield1990salt,winton1993deep,sakai1999dynamical,rial2011modeling}, Earth System Models of Intermediate Complexity (EMICs) \citep{sakai1997dansgaard,wang2006glacial,friedrich2010mechanism}, and coupled GCMs \citep{peltier2014dansgaard}. Many works involve instabilities of the THC (possibly coupled with sea ice) as a mechanism of oscillations, while an instability in atmosphere--ocean--ice--sheet system is also proposed \citep{kageyama2005dansgaard}. 

The third paradigm is related to {\it excitability}. \cite{ganopolski2001rapid} explain DO events by two modes of THC with different convection sites, where the cold mode is stable and the warm mode is marginally unstable. Coherent oscillations between these two modes arise when the freshwater perturbations have an optimal magnitude \citep{ganopolski2002abrupt}. %These two paradigms are reviewed in \citep{crucifix2012oscillators}. %Thus, the difference between excitability and self-sustained oscillations is vague under the noise. 

The FitzHugh-Nagumo model (also known as Bonhoeffer--van der Pol oscillator) is a paradigmatic model, which can exhibit both self-sustained oscillations and excitability \citep{Fitzhugh1961impulses,nagumo1962active}. A particular case of FitzHugh-Nagumo model is introduced as a model of DO events by \cite{kwasniok2013analysis}. Here, we consider a forced version of Kwasniok's model:
\begin{equation}
{\ddot x}+[\alpha +\beta (x-x_*)^2]{\dot x}+k(x-x_0)=\gamma '_0 +\gamma '_1 I(t) -\gamma '_2 V(t).
\end{equation}
where the parameters $x_*$ and $x_0$ control dissipative and linear restoring forces, respectively, and the parameter $\gamma _0'$ is the bias of the external forcing. A priori, $\alpha $ can be either positive or negative. The condition $\alpha <0$ is necessary to exhibit self-sustained oscillations.

By the Li\'enard transformation $v =({\dot x}+\int _ 0^x [\alpha +\beta (x-x_*)^2]dx)/k$
and by adding stochastic forcing terms, we obtain 
\begin{equation}
dx(t) = \left\{kv-\alpha x-\frac{\beta}{3}[(x-x_*)^3+x_*^3]\right\}dt+\sigma _1 dW_1(t), \label{eq:x}
\end{equation}
\begin{equation}
dv(t) = \left\{-x+\gamma _0 +\gamma _1 I(t) -\gamma _2 V(t)\right\}dt+\sigma _2 dW_2(t), \label{eq:y}
\end{equation}
%\begin{equation}
%\begin{pmatrix}
%  dx(t)\\
%  dv(t)\\
%\end{pmatrix}=
%\begin{pmatrix}
%  kv-\alpha x-\frac{\beta}{3}[(x-x_*)^3+x_*^3]\\
%  -x+\gamma _0 +\gamma _1 I(t) -\gamma _2 V(t)\\
%\end{pmatrix}dt+
%\begin{pmatrix}
%  \sigma _1 dW_1(t)\\
%  \sigma _2 dW_2(t)\\
%\end{pmatrix}\label{eq:osci}
%\end{equation}
where $\gamma _0=\gamma _0'/k+x_0$, $\gamma _1=\gamma '_1/k$, and $\gamma _2=\gamma '_2/k$, and $W_1(t)$ and $W_2(t)$ are mutually independent Wiener processes. Again, $x(t)$ is the model's true state for the standardized $-\log _{10}$[Ca$^{2+}$], and the observation model is given by Eq.~(\ref{eq:obs}). In addition to this original model with 10 parameters, we also consider a submodel with 9 parameters obtained by setting $\sigma _1\equiv 0$. We refer to the original model as the oscillator model A and the submodel as the oscillator model B.

\subsection{State and parameter estimation}
The parameter estimation is performed jointly with the state estimation.

\subsubsection{State estimation}
The state estimation is here performed with the Unscented Kalman filter (UKF) \citep{julier2004unscented}. Here, we describe the procedure of the UKF for the oscillator model given by Eqs.~(\ref{eq:x}), (\ref{eq:y}), and (\ref{eq:obs}), but the application for the 1D potential model is straightforward. With the Euler-Maruyama method (\cite{gardinerstochastic}, p.~404), Eqs.~(\ref{eq:x}), (\ref{eq:y}), and (\ref{eq:obs}) can be written in a recursive form:
\begin{equation}
\begin{split}
\mathbf{x}(t+h) &= \mathbf{f}(\mathbf{x}(t),t)+\mathbf{q}(t), \,\,\,\mathbf{q} (t) \sim \mathcal{N}(0,\mathbf{Q}), \\
y(t+h) &= \mathbf{H} \mathbf{x}(t+h)+\eta (t+h),\\
\mathbf{f}(\mathbf{x}(t),t)&=\mathbf{x}(t)+\begin{pmatrix}
  kv-\alpha x-\frac{\beta}{3}[(x-x_*)^3+x_*^3]\\
  -x+\gamma _0 +\gamma _1 I(t) -\gamma _2 V(t)\\
\end{pmatrix}h,
\end{split} \label{eq:ss}
\end{equation}
where $h$ is a time step, $\mathbf{x}(t)=(x(t),v(t))^T\in \mathbf{R}^m$ is a $m$-dimensional state vector ($m=2$ for the oscillator model), $y(t)$ is a scalar observation variable obtained by the operation of observation matrix $\mathbf{H}=(1\,0)$, and $\mathcal{N}(0,\mathbf{Q})$ is the zero mean multivariate normal distribution with covariance matrix 
$$
\mathbf{Q} =
 \begin{pmatrix}
  h\sigma _1^2 & 0\\
  0 & h\sigma _2^2\\
 \end{pmatrix}.
$$

Assume $N$ measurements $y_k =y(t_k)$ ($k=1,\,2,...,N$) obtained at $t_k=t_0+k\Delta T$ in the time interval of $\Delta T$($\geq h$). Denote the measurements up to $k$ by $y_{1:k}=\{y_1, y_2, ...,y_k\}$.
The purpose of {\it filtering} is to obtain the conditional probability density $p(\mathbf{x} (t_k)| y_{1:k})$ given the measurements up to $k$ (so-called {\it filtered density}). 
The Kalman filter provides exact solutions to the filtering problems in particular cases of linear models with Gaussian densities. 
For nonlinear models, the Kalman filter is not available because the filtered density $p(\mathbf{x} (t_k)| y_{1:k})$ deviates from Gaussian. The {\it unscented Kalman filter} (UKF) is one extension of the Kalman filter for nonlinear models, 
where the filtered density $p(\mathbf{x} (t_k)| y_{1:k})$ is approximated by a Gaussian distribution with consideration for the effect of nonlinear transformations \citep{julier2004unscented}. As a result, the filtered density $p(\mathbf{x} (t_k)| y_{1:k})$ is specified only by its mean $\hat{\mathbf{x}}_{k|k}$ (the {\it filtered mean}) and covariance $\mathbf{P}_{k|k}$ (the {\it filtered covariance}).  The details for implementing the UKF are presented, for example, in \cite{kwasniok2012stochastic}, but we repeat them here for self-consistency:

Assume the mean $\hat{\mathbf{x}}_{0|0}$ and covariance $\mathbf{P}_{0|0}$ of the initial density $p(\mathbf{x} (t_0)| y_{1:0})$,
where $y_{1:0}$ is the null set since we have no measurement at $t_0$. The time interval $[t_{k-1},t_k]$ between successive measurements is divided into $L=\Delta T/h$ subintervals.

{\it Prediction step.} Assume that we have calculated the filtered mean and the filtered covariance up to $\hat{\mathbf{x}}_{k-1|k-1}$ and $\mathbf{P}_{k-1|k-1}$. We sequentially calculate a predicted mean $\hat{\mathbf{x}}_l$ and a predicted covariance $\mathbf{P}_l$ at time $t=t_{k-1}+lh$ up to $l=L$, starting with $\hat{\mathbf{x}}_0=\hat{\mathbf{x}}_{k-1|k-1}$ and $\mathbf{P}_0=\mathbf{P}_{k-1|k-1}$. First, we generate an ensemble of state points, which has the same mean and covariance as $\hat{\mathbf{x}}_{l-1}$ and $\mathbf{P}_{l-1}$, the so-called {\it sigma points}:
\begin{eqnarray}
\bm{\chi} ^i_{l-1|l-1}&=&\hat{\mathbf{x}}_{l-1}-(\sqrt{m\mathbf{P}_{l-1}})_i, \nonumber \\
\bm{\chi} ^{i+m}_{l-1|l-1}&=&\hat{\mathbf{x}}_{l-1}+(\sqrt{m\mathbf{P}_{l-1}})_i,\,\,\,i=1,...,m,
\end{eqnarray}
where $(\cdot )_i$ means the $i$th column of a matrix and the square root matrix $\sqrt{m\mathbf{P}_{l-1}}$ is obtained by the Choleski decomposition of $m\mathbf{P}_{l-1}$. Each sigma point $\bm{\chi} ^i_{l-1|l-1}$ is propagated by the deterministic part of Eq.~(\ref{eq:ss}):
$$\bm{\chi} ^i_{l|l-1}=\mathbf{f}(\bm{\chi} ^i_{l-1|l-1},t_k+(l-1)h).$$
Then, the predicted mean $\hat{\mathbf{x}}_l$ and the predicted covariance $\mathbf{P}_l$ at time $t=t_{k-1}+lh$ are given as
$$\hat{\mathbf{x}}_{l}=\frac{1}{2m}\sum _{i=1}^{2m}\bm{\chi} ^i_{l|l-1},\,\,\,
\mathbf{P}_{l}=\frac{1}{2m}\sum _{i=1}^{2m}(\bm{\chi} ^i_{l|l-1}-\hat{\mathbf{x}}_{l})(\bm{\chi} ^i_{l|l-1}-\hat{\mathbf{x}}_{l})^T+\mathbf{Q}.$$
If we reach $l=L$, we go to the following filtering step setting $\hat{\mathbf{x}}_{k|k-1}=\hat{\mathbf{x}}_{L}$ and $\mathbf{P}_{k|k-1}=\mathbf{P}_{L}$.

{\it Filtering step.} The filtered mean $\hat{\mathbf{x}}_{k|k}$ and the filtered covariance $\mathbf{P}_{k|k}$ are given by the usual Kalman filter equation:
\begin{equation}
\begin{split}
\hat{\mathbf{x}}_{k|k} &=\hat{\mathbf{x}}_{k|k-1}+\mathbf{K}_k \mathbf{\zeta }_k,\\
\mathbf{P}_{k|k}       &=(\mathbf{I}-\mathbf{K}_k \mathbf{H})\mathbf{P}_{k|k-1},\\
\mathbf{\zeta }_k&=y_k -\mathbf{H}\hat{\mathbf{x}}_{k|k-1},\\
\mathbf{K}_k &=\mathbf{P}_{k|k-1}\mathbf{H}^{T}S_k^{-1},\\
S_k&=\mathbf{H}\mathbf{P}_{k|k-1}\mathbf{H}^{T}+\varepsilon ^2,
\end{split}
\end{equation}
where $\mathbf{\zeta }_k$ is called the {\it innovation}, $\mathbf{K}_k$ is the Kalman gain matrix, and $S_k$ is the innovation covariance. The prediction and filtering are continued until $k=N$. 

\subsubsection{Parameter estimation}
The likelihood is the conditional probability density of the observations $y_{1:N}$ when the model parameters $\theta $ are given:
$L(\theta ) = p(y_{1:N}|\theta )=\prod _{k=1}^{N} p(y_k|y_{1:k-1},\theta ).$
Owing to the Markov property of Eq.~(\ref{eq:ss}) and the Gaussian approximation of the filtered density in the UKF, the log-likelihood $\ln L(\theta)$ can be written as
\begin{equation}
\ln L(\theta ) = -\frac{N}{2}\ln 2\pi -\frac{1}{2} \sum _{k=1}^N \left( \ln S_k+\frac{\zeta _k^2}{S_k} \right). \label{eq:lnlike}
\end{equation}
The maximum likelihood estimator (MLE) is the parameter that maximizes Eq.~(\ref{eq:lnlike}):
$$\hat{\theta}  =\text{argmax} _{\theta } L(\theta ) =\text{argmax} _{\theta} \ln L(\theta ).$$
The covariance matrix of the MLE is estimated from the hessian matrix of the log-likelihood function:
$$\text{var}(\hat{\theta} )=-\left[ \frac{\partial ^2\ln L(\theta)}{\partial \theta \partial \theta ^T}\right]_{\theta =\hat{\theta} }^{-1}.$$  

We maximize the likelihood $\ln L(\theta )$ by a quasi-Newton method, called the Limited-memory Broyden-Fletcher-Goldfarb-Shanno (L-BFGS-B) method \citep{byrd1994representations} implemented in the R-function optim \citep{team2014r}.  The L-BFGS-B method allows physically-reasonable constraints on the parameter regions (such as $a_4>0$ and $\beta >0$). This procedure may be interpreted as the implementation of bounded uniform priors in a Bayesian framework. To find the global maximum of $\ln L(\theta )$ (not just a local one), we maximize $\ln L(\theta )$ starting from several different values of $\theta$ sampled from an enough wide parameter region.

We note that sequential Monte-Carlo algorithms exist to estimate the likelihood without the possible biases introduced by the Gaussian approximations used in the UKF \citep{andrieu2010particle,chopin2013smc2}. In a separate study, \cite{carson2015bayesian} show that such algorithms may even be implemented to estimate model evidence (Bayes factors, see Section 2.4) in paleoclimate applications, but such algorithms are more computationally expensive. We leave a more systematic comparison of the different algorithms for a future study.

\subsection{Model comparison method}\label{sec:BIC}
The models are compared using several diagnostics: the probability density, the sample autocorrelation function \citep{venables2013modern}, the occurrence frequency of DO warmings $n(t)$ defined in Section~1, and the Bayesian Information Criterion (BIC). Here, we outline the BIC.

In Bayesian model selection, the Bayes factor $B_{ij}$ is used as a standard measure to quantify the evidence in favor of a model (say $M_i$) against model (say $M_j$): $$B_{ij}=\frac{p(y_{1:N}|M_i)}{p(y_{1:N}|M_j)},$$ where the models need not be nested \citep{kass1995bayes}. The quantity $p(y_{1:N}|M_i)$ is the probability density of the data $y_{1:N}$ given a model $M_i$ and obtained as
\begin{equation}
p(y_{1:N}|M_i)=\int p(y_{1:N}|\theta _i,\,M_i)p(\theta _i|M_i)d\theta _i, \label{eq:int}
\end{equation}
where $p(y_{1:N}|\theta _i,\,M_i)$ is the likelihood under the model $M_i$ and $p(\theta _i|M_i)$ is the prior density. 
A value of $B_{ij}>1$ means that the model $M_i$ is more strongly supported by the data than the model $M_j$. 
In many practical cases, the calculation of Eq.~(\ref{eq:int}) is computationally expensive, and reasonable specifications of prior density $p(\theta _i|M_i)$ are difficult. 

One approach for evaluating Bayes factors is the Bayesian Information Criterion (BIC) \citep{kass1995bayes}:
\begin{equation}
\text{BIC}=-2\ln L(\hat{\theta})+K\ln N, \label{eq:BIC}
\end{equation}
where $K$ is the number of the model parameters, $N$ is the number of data points, and $\ln L(\hat{\theta} )$ is the maximum log-likelihood. Denote the Bayesian Information Criterion for model $i$ by $\text{BIC}_i$.
The BIC provides a useful approximation to the Bayes factor for large $N$ with a relative error of $O(N^{-1/2})$:
\begin{equation}
2\ln B_{ij}\approx \text{BIC}_j-\text{BIC}_i=\Delta \text{BIC}_{ij},\label{eq:BFBIC}
\end{equation}
if we assume the {\it unit information prior} on the parameters \citep{kass1995bayes}. The unit information prior is a multivariate normal prior with mean at the maximum likelihood estimator and covariance equal to the expected information matrix for one observation. This can be thought of as an uninformative prior which contains the same amount of information as a single, typical observation \citep{raftery1995bayesian}. Among several models, the model with lowest BIC is preferred. The Akaike Information Criterion, $\text{AIC}=-2\ln L(\hat{\theta})+2K,$ is another popular criterion for model selection \citep{akaike1974new}. The BIC penalizes the number of parameters more strongly than the AIC for large $N$.

\cite{raftery1995bayesian} provides a rule of thumb for interpreting the BIC difference: the evidence is said {\it weak} if $0<\Delta \text{BIC}<2$, {\it positive} if $2<\Delta \text{BIC}<6$, {\it strong} if $6<\Delta \text{BIC}<10$, and {\it very strong} if $\Delta \text{BIC}>10$. In most cases, the BIC difference is more conservative than the Bayes factors based on more informative priors  \citep{raftery1999bayes}; That is, if the BIC difference shows evidence, the Bayes factors based on more informative priors are likely to show evidence. Thus, \cite{raftery1999bayes} recommends to use the BIC difference as a baseline reference quantity. In this study, we have little prior knowledge on parameters, and hence report only the BIC difference as an element of model evidence. 

\section{Results: parameter estimation}
The 1D potential models (A and B) and the oscillator models (A and B) are respectively studied for four cases: no forcing ($\gamma _1\equiv \gamma _2\equiv  0$), insolation forcing ($\gamma _2\equiv 0$), ice volume forcing ($\gamma _1\equiv 0$), and full forcing (i.e., both of insolation and ice volume forcing). We calibrate the models by maximizing the log-likelihood in Eq.~(\ref{eq:lnlike}) with the 20-year average NGRIP record ($-\log _{10}$[Ca$^{2+}$] or $\delta ^{18}$O) in 26--90~ka~BP. To find the global maximum of $\ln L(\theta )$, we maximize $\ln L(\theta )$ starting from 12 different values of $\theta$ randomly sampled from an enough wide parameter region. In the UKF, the time step of $h=0.001$~ka is used for efficient and robust estimations of MLEs. The initial conditions for the UKF are set as $\hat{\mathbf{x}}_{0|0}=y_1$ (the value at 90~ka~BP) and $\mathbf{P}_{0|0}=\varepsilon ^2$ for the 1D potential models and as $\hat{\mathbf{x}}_{0|0}=(y_1,\,y_1)^T$ and $\mathbf{P}_{0|0}=\text{diag}(\varepsilon ^2,\,10)$ for the oscillator models, where $\varepsilon$ is chosen from $10^{-6}<\varepsilon <0.2$ for the first calculation of $\ln L(\theta )$ and then updated in the optimization process. It takes several time-steps before the influence of the initial conditions on the filtered densities vanishes. Therefore, the first 50 elements are excluded from the summation in the log-likelihood $\ln L(\theta )$ in Eq.~(\ref{eq:lnlike}) in order to effectively discard the influence of initial conditions. Then, the data length becomes $N=3151$. Also in numerical integrations, we use the Euler-Maruyama method with the time step of $h=0.001$~ka.

\subsection{Maximum likelihood estimate for the 1D potential model: the case of Ca$^{2+}$ record}
\begin{figure*}
\begin{center}
  \includegraphics[angle=270,width=0.8\linewidth]{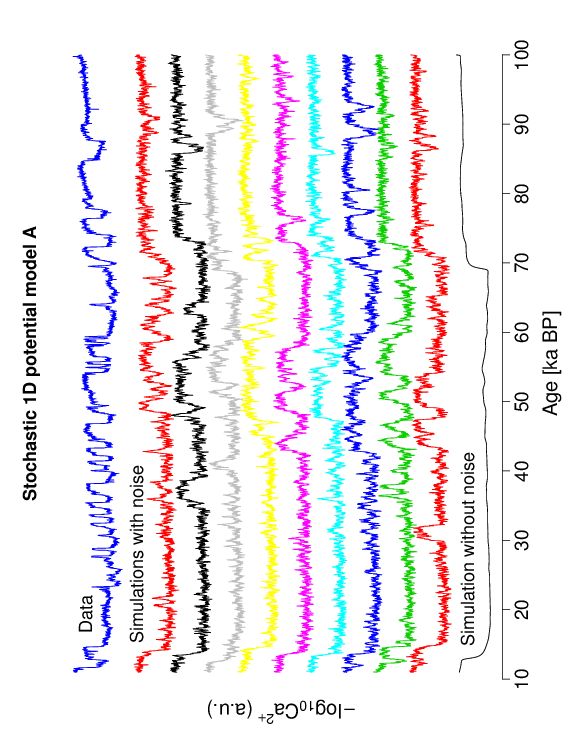}
  \includegraphics[angle=270,width=0.8\linewidth]{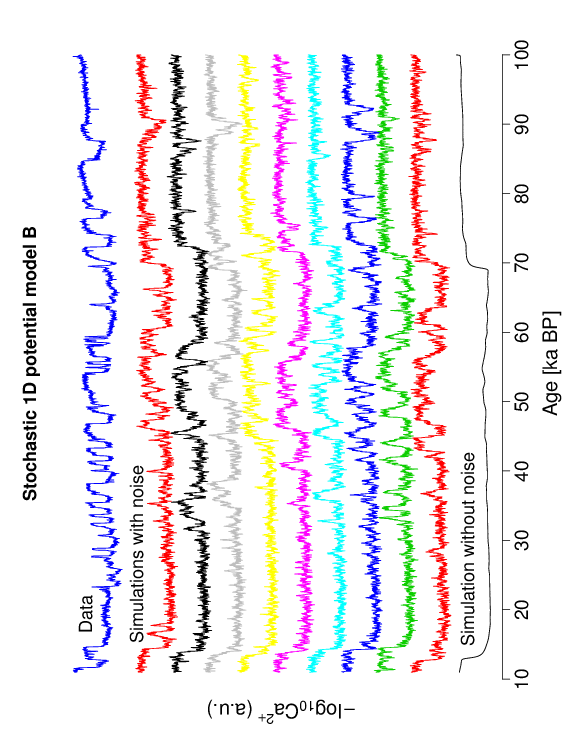}
  \caption{Sample trajectories simulated by the 1D potential model A (top) and model B (bottom) under the full forcing (9 with noise and 1 without noise, i.e.,  $\sigma =\varepsilon =0$).} \label{fig:traj1}
\end{center}
\end{figure*}

For the 1D potential models (A and B) with different forcings, the log-likelihood is maximized with the 20-year average NGRIP Ca$^{2+}$ record. We found a unique maximum for $\ln L(\theta )$ in either case.
%\footnote{The log-likelihood was maximized from 12 different initial conditions of parameter $\theta$. We checked the robustness of the result by changing the intervals, from which initial values of parameters are taken.} 
Tables~\ref{tb:1D1} and \ref{tb:1D2} show the maximum likelihood estimator $\hat{\theta }$, the maximum log-likelihood $\ln L(\hat{\theta })$, the BIC and the AIC. Based on the BIC, the full forcing is preferred in both models A and B. The same conclusion is obtained if the AIC is used. Sample trajectories of the fully forced models corresponding to different noise realizations are shown in Fig.~\ref{fig:traj1}. 
%The long-term trends are similar as the observed record, but the transitions are less frequent. 
The lowest BIC of model B is slightly lower than that of model A, but the difference is less than 2. Thus, selecting model A or model B is difficult. In other words, the contribution of the observation noise is uncertain. However, it should be noted that the inference on the forcing is robust regardless of the uncertainty in the noise.

\begin{figure}
\begin{center}
  \includegraphics[angle=270,width=0.9\linewidth]{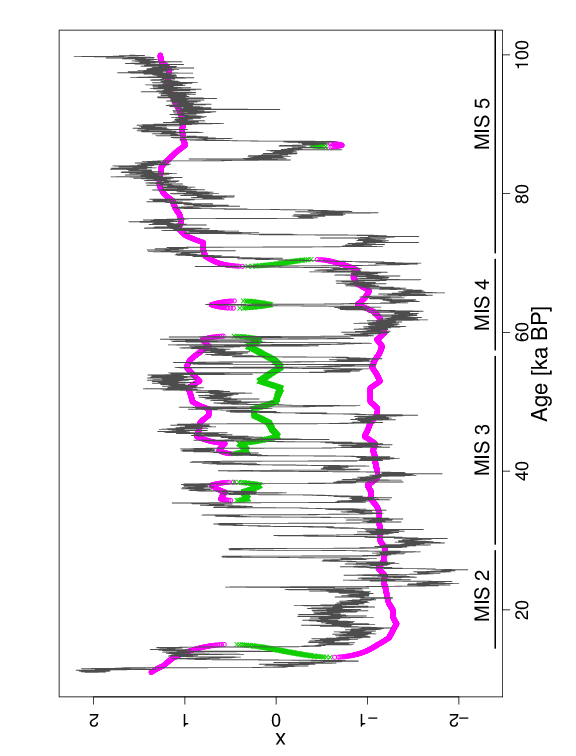}
  \caption{Locations of the local minima (magenta, circle) and the local maximum (green, cross) of the effective potential $U_{\text{eff}}(x,t)$ with respect to $x$. The case of the 1D potential model A with full forcing (see Table~\ref{tb:1D1} for the parameters). Roughly speaking, the state $x(t)$ is stable near the local minima and unstable near the local maximum. The observational data $y$ is represented by a line.} \label{fig:equilibria}
\end{center}
\end{figure}

The stability of the system is grasped by the effective potential $U_{\text{eff}}(x,t)=U(x)-x[\gamma _1I(t) -\gamma _2 V(t)]$. Roughly speaking, the state $x(t)$ is stable near the local minima of $U_{\text{eff}}(x,t)$ with respect to $x$ and unstable near the local maximum. Though $U(x)$ has always two local minima, $U_{\text{eff}}(x,t)$ has either single or two local minima depeding on the time-varying forcings as shown in Fig.~\ref{fig:equilibria}. Temporal changes of $U_{\text{eff}}(x,t)$ are almost the same in the model A and B. The MIS~5 is characterized mainly by a monostable interstadial state. Due to the decreased insolation $I(t)$ and the increasing ice volume $V(t)$, the interstadial state looses stability, and a stable stadial state appears in the MIS~4. In the early part of MIS~3, the system becomes bistable due to the increased insolation $I(t)$. In the late part of the MIS~3, the system goes gack to the monostable stadial state until the deglaciation. These stability changes are qualitatively similar with the result of nonlinear potential analysis by \cite{livina2010potential} and the EMIC simulation by \cite{ganopolski2001rapid}, which shows the existence of a stable stadial state and a marginally unstable interstadial state in a glacial condition. %, and a fully coupled GCM simulation \citep{zhang2014abrupt}, which shows the appearance of bistablity for an intermediate size of ice sheet.   

\begin{landscape}
%\begin{table}
%\begin{tabular}{lll}
%\hline\noalign{\smallskip}
%\end{tabular}
%\end{table}
\begin{table}
\caption{Maximum likelihood estimate for the 1D potential model A (the NGRIP $-\log _{10}[\text{Ca}^{2+}]$ data). For each parameter, the maximum likelihood estimator is shown with the standard error.} \label{tb:1D1}
\scalebox{0.6}{
\begin{tabular}{ l  c  c  c  c  c  c  c  c  c  c  c  c  c}
\hline\noalign{\smallskip}
\textbf{1D model A} & $a_1$ & $a_2$ & $a_3$ & $a_4$ & $\sigma $ & $\varepsilon $ & $\gamma _1$ & $\gamma _2$ & $\ln L$ & BIC & AIC & $K$\\ 
\noalign{\smallskip}\hline\noalign{\smallskip}
Full forcing    & $0.048\pm 0.306$  & $-1.6\pm 0.2$ & $0.14\pm 0.09$  & $1.0\pm 0.1$  & $1.39\pm 0.04$ & $0.046\pm 0.009$ & $0.72\pm 0.20$ & $2.4\pm 0.3$ & $719.4$ & $-1374.4$ & $-1422.8$ & 8\\ 
Insolation only & $0.31\pm 0.29$   & $-1.6\pm 0.2$ & $0.012\pm 0.089$ & $0.83\pm 0.08$ & $1.34\pm 0.04$ & $0.054\pm 0.007$ & $0.75\pm 0.19$ & $\equiv 0$ & $691.3$ & $-1326.2$ & $-1368.6$ & 7 \\
Ice volume only & $-0.081\pm 0.303$ & $-1.7\pm 0.2$ & $0.15\pm 0.09$  & $1.0\pm 0.09$   & $1.38\pm 0.03$ & $0.048\pm 0.008$  & $\equiv 0$ & $2.5\pm 0.3$ & $712.4$ & $-1368.5$ & $-1410.8$ & 7\\ 
No forcing     & $0.19\pm 0.29$   & $-1.7\pm 0.2$ & $0.018\pm 0.088$ & $0.80\pm 0.08$ & $1.33\pm 0.03$   & $0.056\pm 0.007$  & $\equiv 0$ & $\equiv 0$ & $683.3$ & $-1318.2$ & $-1354.6$ & 6\\
\noalign{\smallskip}\hline
\end{tabular}
}
\end{table}

\begin{table}
\caption{Maximum likelihood estimate for the 1D potential model B (the NGRIP $-\log _{10}[\text{Ca}^{2+}]$ data). } \label{tb:1D2}
\scalebox{0.6}{
\begin{tabular}{ l  c  c  c  c  c  c  c  c  c  c  c  c}
\hline\noalign{\smallskip}
\textbf{1D model B} & $a_1$ & $a_2$ & $a_3$ & $a_4$ & $\sigma $ & $\varepsilon $ & $\gamma _1$ & $\gamma _2$ & $\ln L$ & BIC & AIC & $K$\\ 
\noalign{\smallskip}\hline\noalign{\smallskip}
Full forcing    & $0.062\pm 0.325$  & $-1.8\pm 0.25$ & $0.15\pm 0.10$  & $1.2\pm 0.09$ & $1.48\pm 0.02$ & $\equiv 0$  & $0.80\pm 0.21$ & $2.7\pm 0.3$ & $716.1$ & $-1375.8$ & $-1416.1$ & 7\\ 
Insolation only & $0.37\pm 0.32$   & $-1.9\pm 0.2$ & $0.006\pm 0.096$ & $0.97\pm 0.08$ & $1.46\pm 0.02$ & $\equiv 0$ & $0.87\pm 0.20$ & $\equiv 0$ & $684.3$ & $-1320.2$ & $-1352.4$ & 6\\
Ice volume only & $-0.073\pm 0.322$ & $-1.9\pm 0.2$ & $0.16\pm 0.10$  & $1.1\pm 0.09$ & $1.47\pm 0.02$ & $\equiv 0$ & $\equiv 0$ & $2.7\pm 0.3$ & $708.5$ & $-1368.5$ & $-1400.9$ & 6\\ 
No forcing     & $0.22\pm 0.31$   & $-2.0\pm 0.2$ & $0.012\pm 0.095$ & $0.94\pm 0.08$ & $1.46\pm 0.02$ & $\equiv 0$ & $\equiv 0$ & $\equiv 0$ & $675.2$ & $-1310.1$ & $-1334.2$ & 5\\
\noalign{\smallskip}\hline
\end{tabular}
}
\end{table}

\begin{table}
\caption{Maximum likelihood estimate for the oscillator model A (the NGRIP $-\log _{10}[\text{Ca}^{2+}]$ data).} \label{tb:mode1}
\scalebox{0.6}{
\begin{tabular}{  l  c  c  c  c  c  c  c  c  c  c  c  c  c  c  c c}
\hline\noalign{\smallskip}
\textbf{Oscillator model A}  & $k$ & $x_*$ & $\alpha $ & $\beta $ & $\sigma _1$ & $\sigma _2$ & $\varepsilon$ & $\gamma _0$ & $\gamma _1$ & $\gamma _2$ & $\ln L$  & BIC & AIC & $K$\\ 
\noalign{\smallskip}\hline\noalign{\smallskip}
  Full forcing             & $38\pm 8$ & $-0.081\pm 0.017$ & $-0.32\pm 2.00$ &  $58\pm 6$ &  $1.3\pm 0.1$ & $0.78\pm 0.11$ & $0.074\pm 0.010$ & $-0.15\pm 0.10$ & $0.34\pm 0.10$ &  $0.57\pm 0.14$ & $914.8$ & $-1748.9$ & $-1809.6$ & 10\\ 
  Insolation only          & $28\pm 6$ & $-0.079\pm 0.017$ & $-0.17\pm 2.05$ & $59\pm 6$ & $1.3\pm 0.1$ & $1.1\pm 0.2$ & $0.073\pm 0.010$ & $-0.19\pm 0.14$ & $0.48\pm 0.13$ & $\equiv 0 $ & $909.2$ & $-1745.9$ & $-1800.4$ & 9\\  
  Ice volume only          & $33\pm 7$ & $-0.080\pm 0.017$ & $-0.098\pm 2.067$ & $59\pm 6$ &  $1.3\pm 0.1$ & $0.93\pm 0.15$ & $0.075\pm 0.010$ & $-0.13\pm 0.12$ & $\equiv 0 $ & $0.70\pm 0.16$ & $910.0$ & $-1747.5$ & $-1802.0$ &9  \\ 
  No forcing               & $22\pm 5$ & $-0.078\pm 0.017$ & $0.013\pm 2.160$ & $60\pm 6$ &  $1.3\pm 0.1$ & $1.4\pm 0.3$ & $0.073\pm 0.010$ & $-0.16\pm 0.18$ & $\equiv 0 $ & $\equiv 0 $ & $904.1$ & $-1743.7$ & $-1792.2$ & 8 \\  
\noalign{\smallskip}\hline
\end{tabular}
}
\end{table}

\begin{table}
\caption{Maximum likelihood estimate for the oscillator model B (the NGRIP $-\log _{10}[\text{Ca}^{2+}]$ data).} \label{tb:mode2}
\scalebox{0.6}{
\begin{tabular}{  l  c  c  c  c  c  c  c  c  c  c  c  c  c  c c}
\hline\noalign{\smallskip}
  \textbf{Oscillator model B} & $k$ & $x_*$ & $\alpha $ & $\beta $ & $\sigma _2$ & $\varepsilon$ & $\gamma _0$ & $\gamma _1$ & $\gamma _2$ & $\ln L$ & BIC & AIC & $K$\\ 
\noalign{\smallskip}\hline\noalign{\smallskip}
               Full forcing & $100\pm 17$ & $-0.052\pm 0.017$ & $17\pm 5$ & $130\pm 13$ &  $0.88\pm 0.14$ &  $0.12\pm 0.003$ &  $-0.15\pm 0.11$ &  $0.33\pm 0.11$ &  $0.57\pm 0.16$ & $911.2$ & $-1749.9$ & $-1804.4$ & 9\\ 
            Insolation only &  $75\pm 15$ &  $-0.051\pm 0.017$ &  $18\pm 5$ & $134\pm 14$ &  $1.2\pm 0.2$ & $0.12\pm 0.003$ &  $-0.19\pm 0.15$ &  $0.48\pm 0.15$ & $\equiv 0 $ & $906.7$ & $-1748.9$ & $-1797.4$ & 8\\  
            Ice volume only & $86\pm 16$ & $-0.053\pm 0.017$ &  $17\pm 5$ & $131\pm 13$ & $1.0\pm 0.2$ &  $0.12\pm 0.002$ & $-0.12\pm 0.13$ &  $\equiv 0 $ & $0.70\pm 0.18$ & $907.4$ & $-1750.4$ & $-1798.8$ & 8\\ 
                No forcing  & $58\pm 13$ & $-0.051\pm 0.017$ & $18\pm 5$ & $136\pm 14$ & $1.6\pm 0.3$ & $0.12\pm 0.002$ & $-0.15\pm 0.20$ & $\equiv 0 $ & $\equiv 0 $ & $902.5$ & $-1748.6$ & $-1791.0$ & 7\\  
\noalign{\smallskip}\hline
\end{tabular}
}
\end{table}

\begin{table}
\caption{Maximum likelihood estimate for the 1D potential model (the NGRIP $\delta ^{18}$O$_{\text{ice}}$ data). } \label{tb:ox1}
\scalebox{0.6}{
\begin{tabular}{ c  c  c  c  c  c  c  c  c  c c}
\hline\noalign{\smallskip}
$a_1$ & $a_2$ & $a_3$ & $a_4$ & $\sigma $ & $\varepsilon $ & $\gamma _1$ & $\gamma _2$ & $\ln L$ \\ 
\noalign{\smallskip}\hline\noalign{\smallskip}
$0.46\pm 0.35$  & $-0.31\pm 0.29$ & $0.011\pm 0.099$  & $0.55\pm 0.08$   & $1.7\pm 0.07$ & $0.27\pm 0.007$ & $1.2\pm 0.3$ & $1.9\pm 0.4$ & $-1578.0$ \\ 
\noalign{\smallskip}\hline
\end{tabular}
}
\end{table}

\begin{table}
\caption{Maximum likelihood estimate for the oscillator model A (the NGRIP $\delta ^{18}$O$_{\text{ice}}$ data). %(Note 2) The standard error of $\sigma _1$ is not well-determined from the Hessian matrix of the log-likelihood function for this mode; it is virtually flat over $0<\sigma _1<0.3$.
} \label{tb:ox2}
\scalebox{0.6}{
\begin{tabular}{  c  c  c  c  c  c  c  c  c  c  c  c  c}
\hline\noalign{\smallskip}
$k$ & $x_*$ & $\alpha $ & $\beta $ & $\sigma _1$ & $\sigma _2$ & $\varepsilon$ & $\gamma _0$ & $\gamma _1$ & $\gamma _2$ & $\ln L$ \\ 
\noalign{\smallskip}\hline\noalign{\smallskip}
$581\pm 131$ & $-0.63\pm 0.06$ & $165\pm 28$ & $74 \pm 14$ & $5.7\pm 0.4$ & $0.62\pm 0.07$& $10^{-6}$ (lower bound) & $-0.13\pm 0.08$ & $0.33\pm 0.08$ & $0.61 \pm 0.11$ & $-1511.4$ \\
\noalign{\smallskip}\hline
\end{tabular}
}
\end{table}
\end{landscape}  

\subsection{Maximum likelihood estimate for the oscillator model: the case of Ca$^{2+}$ record}
\begin{figure*}
\begin{center}
  \includegraphics[angle=270,width=0.8\linewidth]{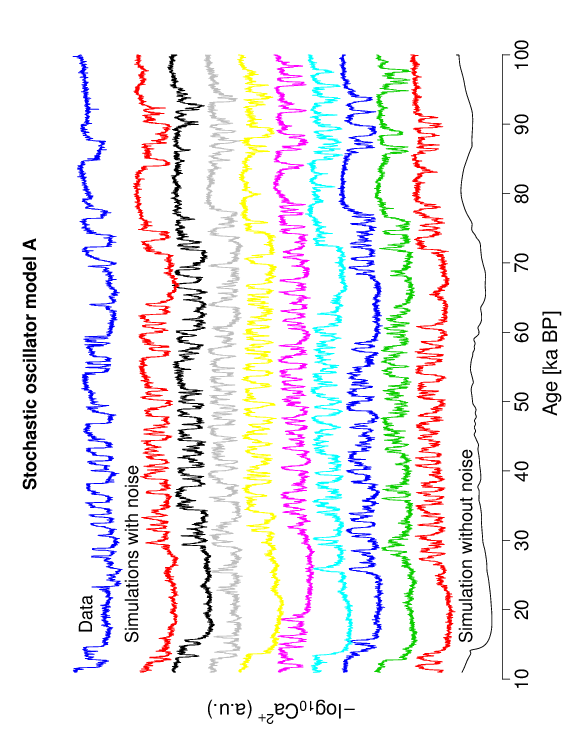}
  \includegraphics[angle=270,width=0.8\linewidth]{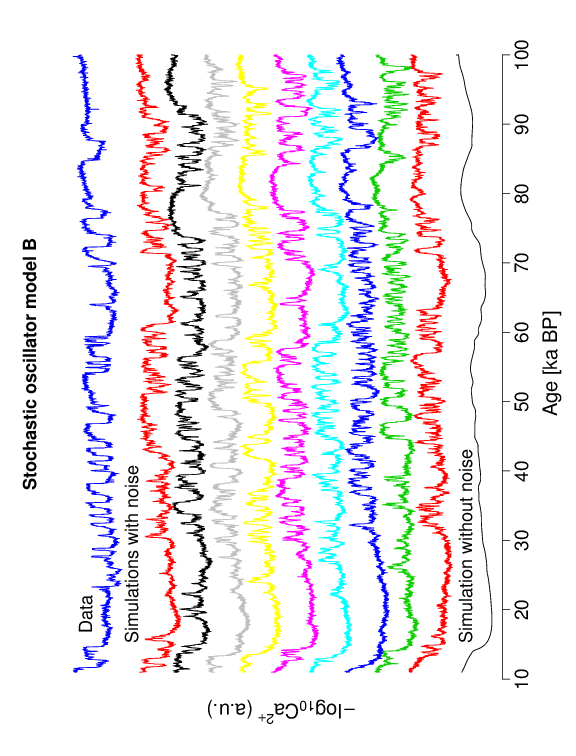}
  \caption{Sample trajectories simulated by the oscillator model A (top) and model B (bottom) under the full forcing (9 with noise and 1 without noise, {\it i.e.},  $\sigma _1 =\sigma _2=\varepsilon =0$).} \label{fig:traj2}
\end{center}
\end{figure*}

The oscillator models A and B with different forcings are calibrated with the 20-year average NGRIP Ca$^{2+}$ record. 
The oscillator model A has two significant local maxima in the likelihood function $L(\theta )$ (see Supplementary Fig.~S1).
The MLE of model A is characterized by a small negative value of $\alpha $ and a relatively large value of $\sigma _1$ (Table \ref{tb:mode1}).
The MLE of model B is characterized by a large positive value of $\alpha $ and $\sigma _1 =0$ (Table \ref{tb:mode2}). This corresponds to the second local maximum of the model A. Because $k,\,\beta \gg 1$ in both models, Eq.~(\ref{eq:x}) is the fast system, and Eq.~(\ref{eq:y}) is the slow system. Note the slow system parameters $\gamma _{0,1,2}$ and $\sigma _2$ have almost the same mean and standard deviation in both models while some of the fast system parameters, $k$, $\alpha$, and $\beta$, are rather different. This suggests that the inference on the external forcings is robust. Based on BIC scores, the models with full forcing or the ice volume forcing are rather preferred to those with insolation forcing or without forcing (in both models A and B). This is also the same if the AIC is used. Figure~\ref{fig:traj2} shows sample trajectories of oscillator model A (top) and B (bottom) under the full forcing, respectively. There seems to be no substantial difference between A and B. Indeed, given that the difference between the lowest BICs is less than 2, it is difficult to select one from A and B. 

\begin{figure}
\begin{center}
  \includegraphics[angle=270,width=0.9\linewidth]{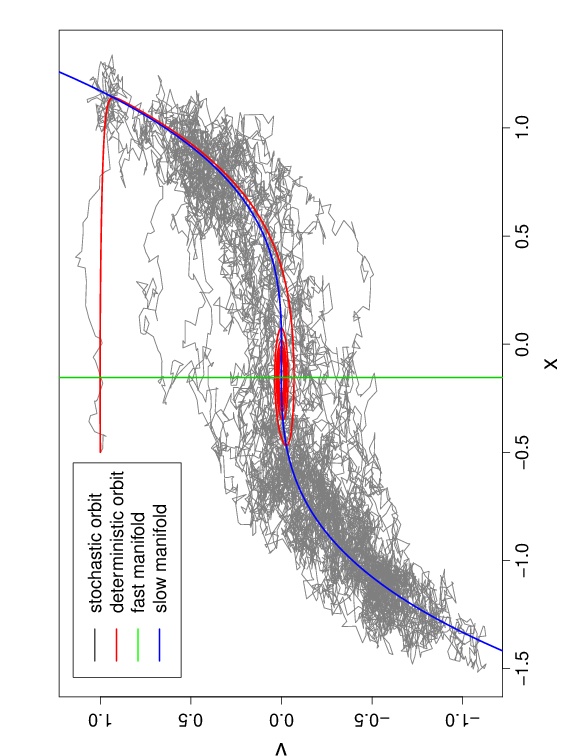}
  \caption{A sample trajectory of the oscillator model A of Eqs.~(\ref{eq:x}) and (\ref{eq:y}) for $\gamma _1 =\gamma _2=0$ (gray). The red curve is a trajectory approaching the limit cycle for the deterministic system for $\sigma _1=\sigma _2=0$ and $\gamma _1 =\gamma _2=0$. The other parameters are set to the maximum likelihood estimator in Table~\ref{tb:mode1} (the full forcing case).} \label{fig:phase_space}
\end{center}
\end{figure}

For better understanding of the dynamics, let us consider the deterministic system obtained by setting $\sigma _1=\sigma _2=0$ and replacing $\gamma _1I(t) -\gamma _2 V(t)$ by a constant external forcing $F_{\text{ext}}$ in Eqs.~(\ref{eq:x}) and (\ref{eq:y}). For the model A (with $\hat{\alpha}=-0.32<0$), the deterministic system has a limit cycle if the external forcing is $-0.025<F_{\text{ext}}<0.171$. However, the limit cycle is much smaller than the observed stochastic cycles in the presence of noise, as shown in Fig.~\ref{fig:phase_space}. The stochastic cycles are formed around the {\it slow manifold} of the system, $v=\{\alpha x+\frac{\beta}{3}[(x-x_*)^3+x_*^3]\}/k$, away from the limit cycle. Hence, they are termed {\it noise-induced oscillations}. This result is consistent with the result obtained by \cite{kwasniok2013analysis} for the unforced case. For the model A, the sign of parameter $\alpha $ is actually uncertain because of the large standard error ($2.0$), but noise-induced oscillations appear regardless of the sign of $\alpha$. For the model B ($\hat{\alpha} >0$), the deterministic system never exhibits self-sustained oscillations, but the system can exhibit {\it noise-induced oscillations}.

\begin{figure}
\begin{center}
  \includegraphics[angle=270,width=0.9\linewidth]{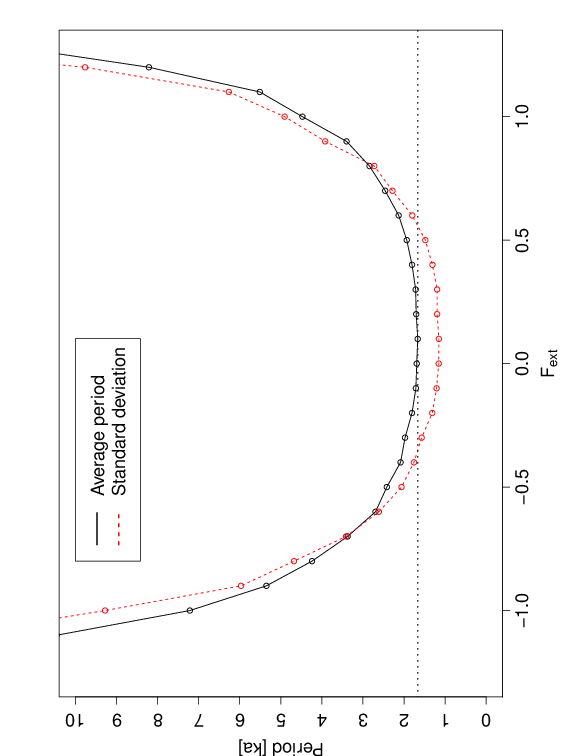}
  \caption{The average period (black, solid) and its standard deviation (red, dashed) between successive warming transitions simulated by stochastic oscillator model A. They are presented as functions of the constant external forcing $F_{\text{ext}}$. See Section~1 for the definition of a warming transition. The average period takes the minimum $\sim$1665~year at $F_{\text{ext}}\approx 0.1$. The parameters are set to the maximum likelihood estimator in Table~\ref{tb:mode1} (the full forcing case).} \label{fig:bias}
\end{center}
\end{figure}

The time scale of noise-induced oscillations emerges from the interplay between the underlying deterministic system and the system noise.\footnote{The average period between successive warming transitions under stochastic noise is not determined by the eigenfrequency of the equilibrium point for the deterministic case. For the case of model A with $F_{\text{ext}}=0$, the former period is $\sim$1700 years but the latter period is $\sim$1000 years.} Figure~\ref{fig:bias} shows the average period between successive warming transitions as a function of the constant external forcing $F_{\text{ext}}$ (for the case of model A), where a warming transition is defined as in Section~1. The U-shape dependency of the period is similar to that of deep-decoupling oscillation models \citep{winton1993deep,schulz2002tempo,colin2007simple}, where the freshwater flux is the control parameter. Using a deep-decoupling model forced by freshwater flux proportional to a reconstructed ice volume, \cite{sima2004younger} argue that the Younger Dryas event may be an intrinsic feature associated with deglaciations, and it seems to be not an accident but an inevitable one. In our ensemble simulations in Fig.~\ref{fig:traj2}, the occurrence of Younger Dryas-type event depends on the realization of system noise.  

\subsection{Maximum likelihood estimate: the case of $\delta ^{18}$O$_{\text{ice}}$ record}
The 1D potential model A and the oscillator model A are calibrated on the 20-year average NGRIP $\delta ^{18}$O$_{\text{ice}}$ record \citep{rasmussen2014stratigraphic} to assess proxy dependence. As shown in Tables~\ref{tb:1D1}--\ref{tb:ox2}, the values of forcing parameters $\gamma _{1,2}$ are consistent between Ca$^{2+}$ and $\delta ^{18}$O$_{\text{ice}}$, but the other parameters are rather different. On the other hand, the values of maximum log-likelihood for $\delta ^{18}$O$_{\text{ice}}$ ($-1578.0$ and $-1511.4$) are significantly lower than those for Ca$^{2+}$ ($719.4$ and $914.8$). This may be taken as an informal indicator that the fit on $\delta ^{18}$O$_{\text{ice}}$ is poorer than on Ca$^{2+}$. Therefore, in this study, we discuss the influence of external forcings based on the models calibrated on Ca$^{2+}$.

\section{Model comparisons}
The models calibrated in the previous Section are compared using several criteria. The probability density is a useful criterion for assessing stochastic dynamical systems models, but here all models reproduce the probability density of the record relatively well. Hence, it is only discussed in the Supplementary Fig.~S2.  

\subsection{Sample autocorrelation function}
\begin{figure*}
\begin{center}
  \includegraphics[angle=270,width=0.4\linewidth]{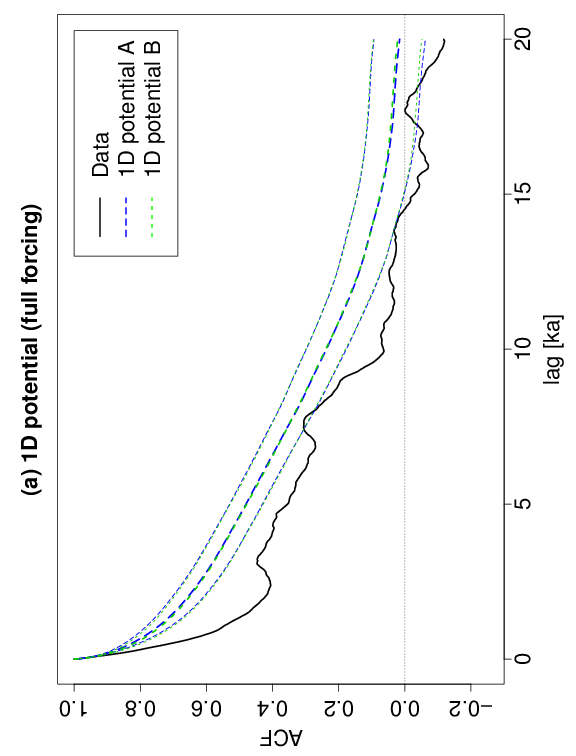}
  \includegraphics[angle=270,width=0.4\linewidth]{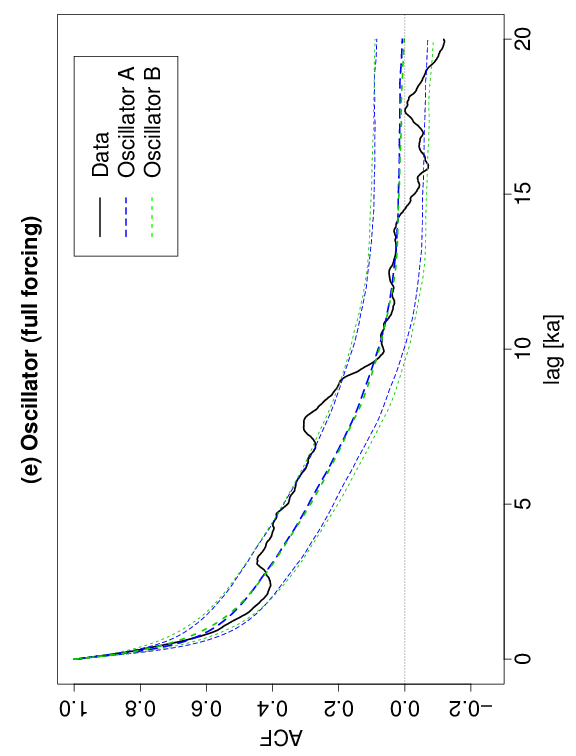}
  \includegraphics[angle=270,width=0.4\linewidth]{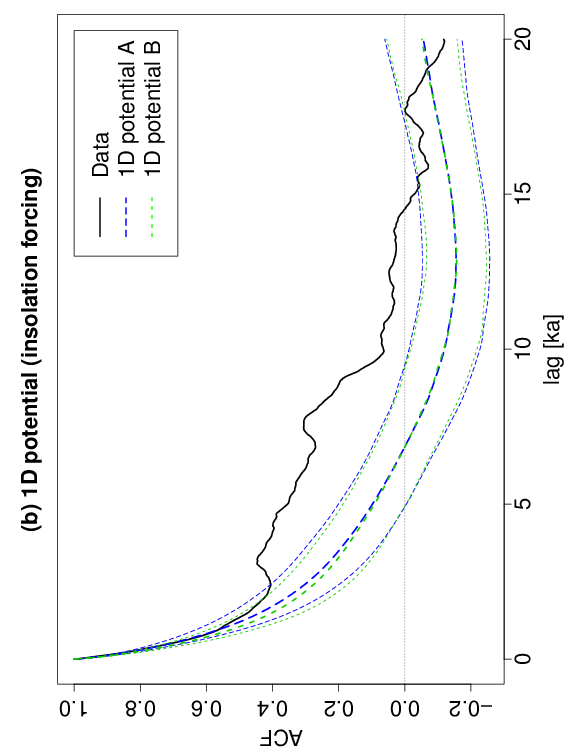}
  \includegraphics[angle=270,width=0.4\linewidth]{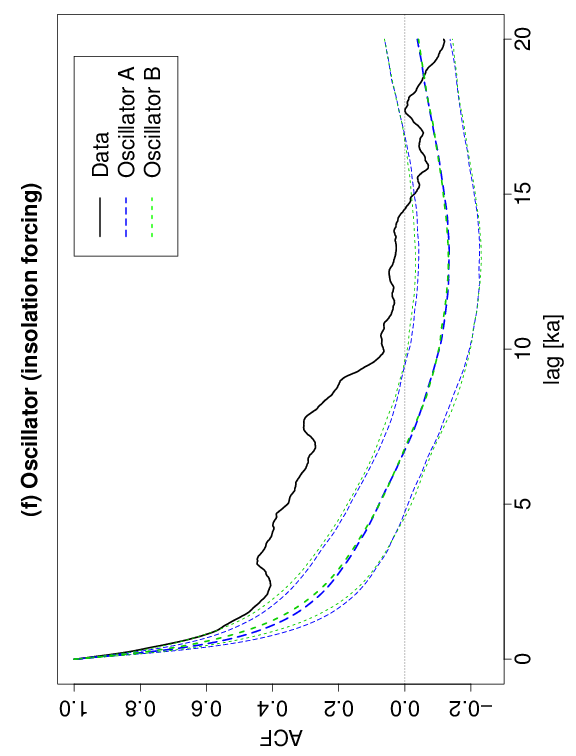}
  \includegraphics[angle=270,width=0.4\linewidth]{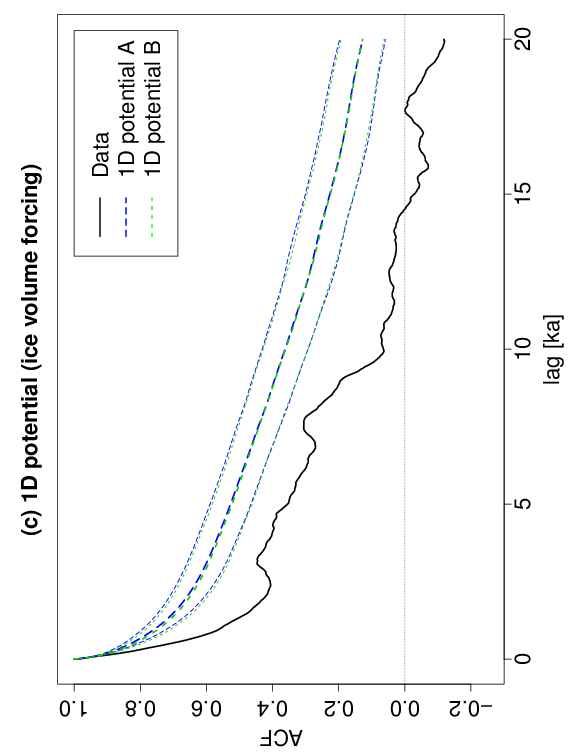}
  \includegraphics[angle=270,width=0.4\linewidth]{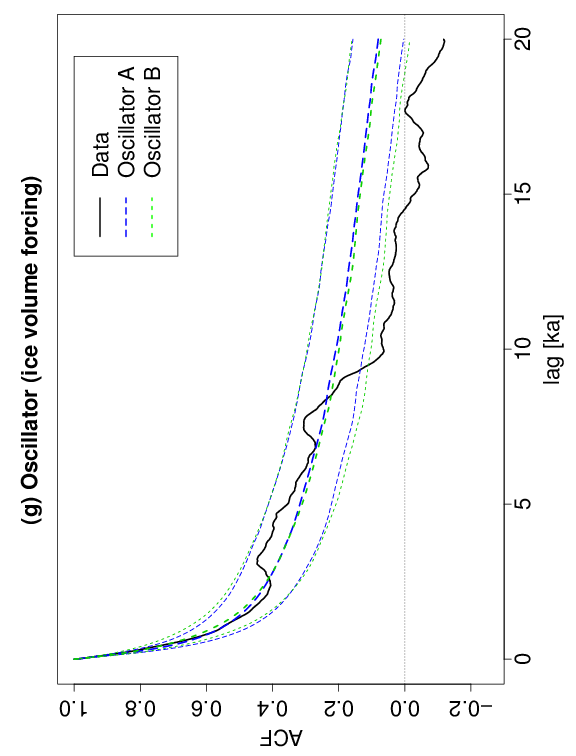}
  \includegraphics[angle=270,width=0.4\linewidth]{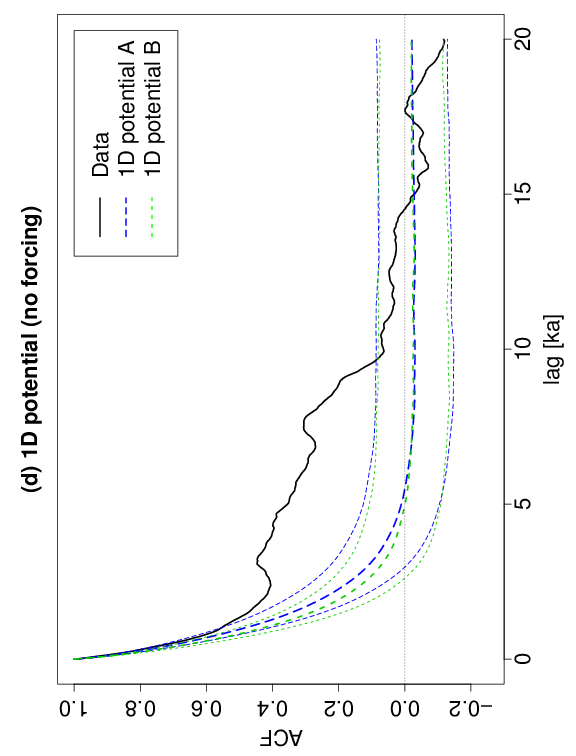}
  \includegraphics[angle=270,width=0.4\linewidth]{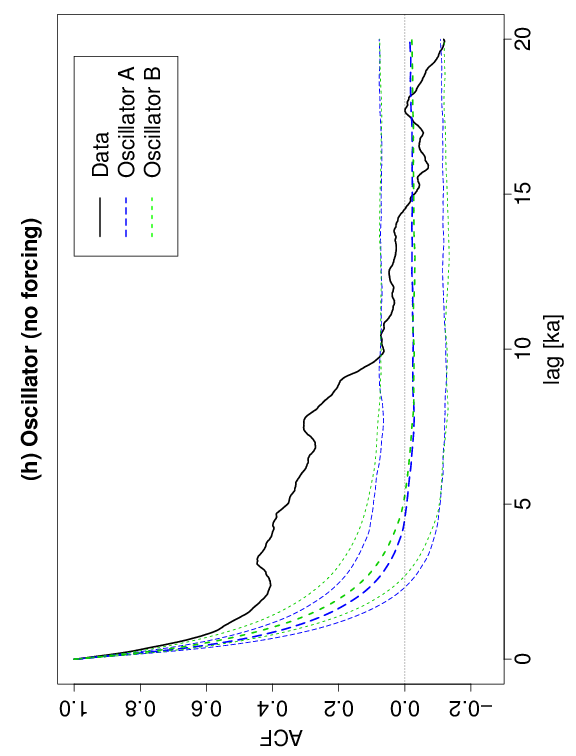}
  \caption{Sample autocorrelation function (ACF) of the 20-year average NGRIP Ca$^{2+}$ data (solid line) and the ensemble mean of the sample ACF of each model (inner dashed line). The outer dashed lines show $\pm 1$ s.d. All sample ACFs are calculated over 11--100~ka BP. Initial conditions are randomly taken at $100$~ka~BP from a normal distribution $p(x)=\mathcal{N} (1,1)$ for the 1D potential model, and from $p(x)= \mathcal{N}(1,1)$ and $p(v)=\mathcal{N} (1,1)$ for the stochastic oscillator models. The ensemble size is $10^3$ for all the models. The ensemble mean of the sample ACFs does not converge to zero even in the stationary cases (d) and (h) because of the finite sample length effect \citep{trenberth1984some}.} \label{fig:acf}
\end{center}
\end{figure*}

First, we assess the model performance by the sample autocorrelation function (ACF) as in \cite{kwasniok2009deriving}. Figure~(\ref{fig:acf}) shows the sample ACF of the 20-year average NGRIP Ca$^{2+}$ data (solid line) and the ensemble mean of the sample ACF simulated by each model (inner dashed line). The outer dashed lines present $\pm 1$ s.d. The sample ACF of the data is explained well by the oscillator models with the full forcing and relatively well by the oscillator models with the ice volume forcing. The difference between the model A and B is negligible.  

\subsection{Model comparison based on the occurrence frequency of warming transitions}
\begin{figure*}
\begin{center}
  \includegraphics[angle=270,width=0.4\linewidth]{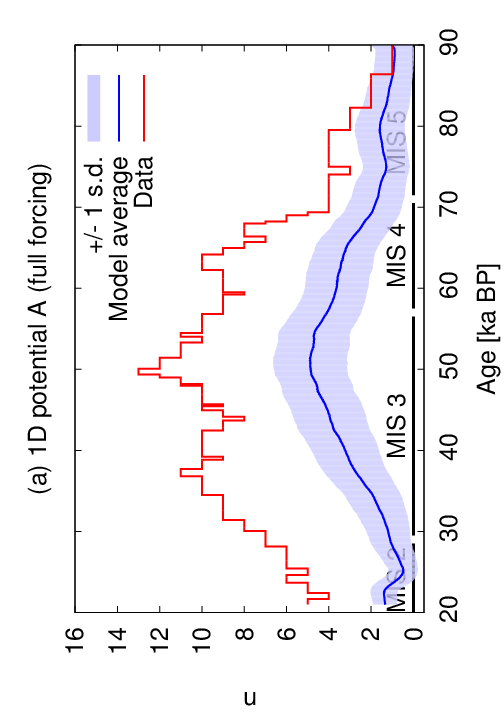}
  \includegraphics[angle=270,width=0.4\linewidth]{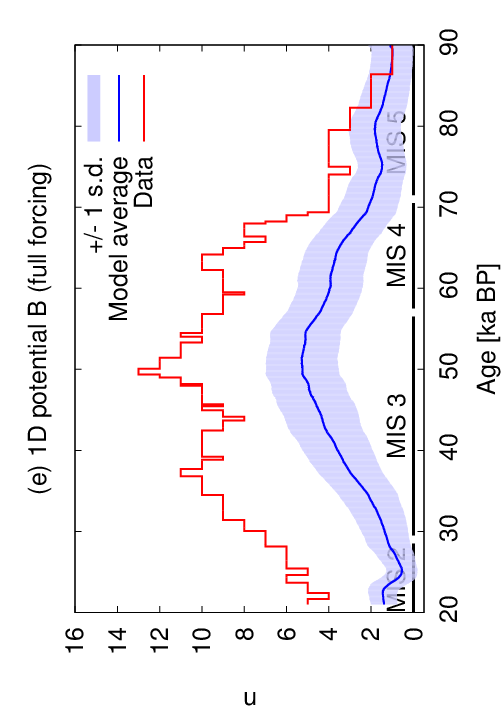}
  \includegraphics[angle=270,width=0.4\linewidth]{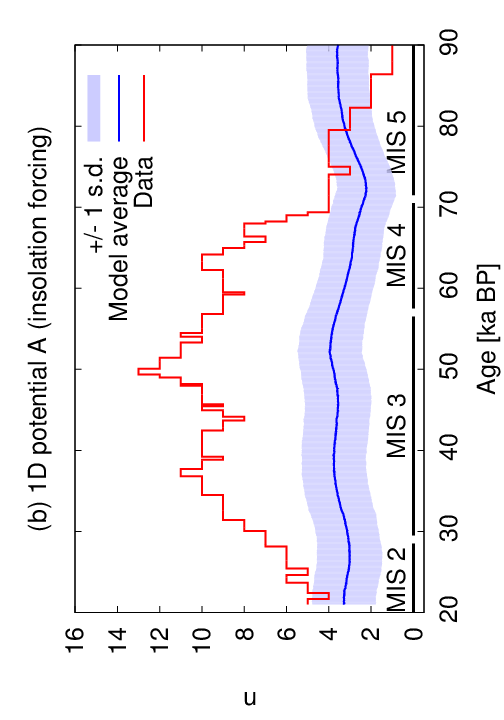}
  \includegraphics[angle=270,width=0.4\linewidth]{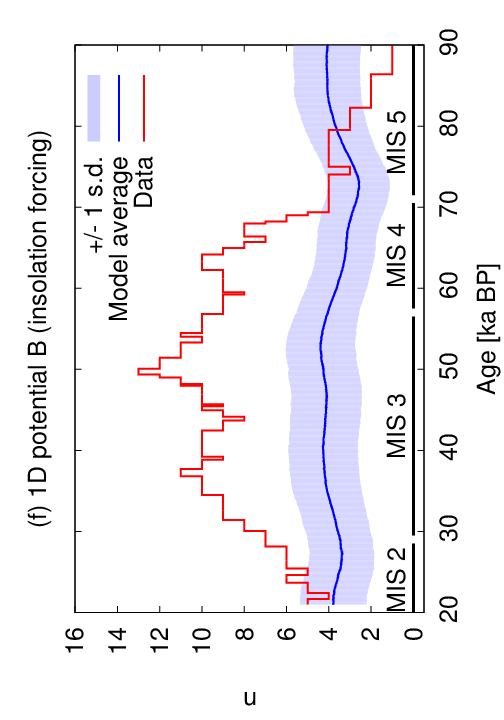}
  \includegraphics[angle=270,width=0.4\linewidth]{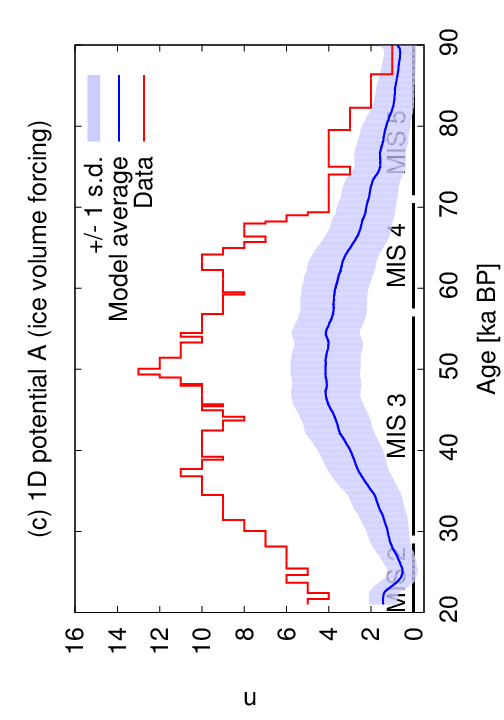}
  \includegraphics[angle=270,width=0.4\linewidth]{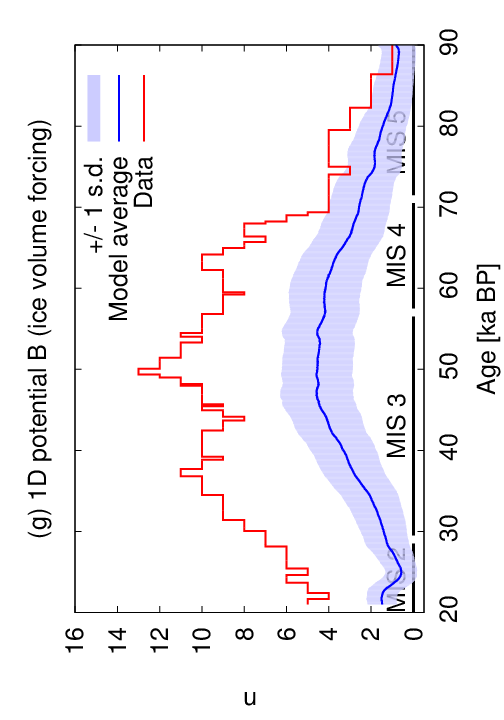}
  \includegraphics[angle=270,width=0.4\linewidth]{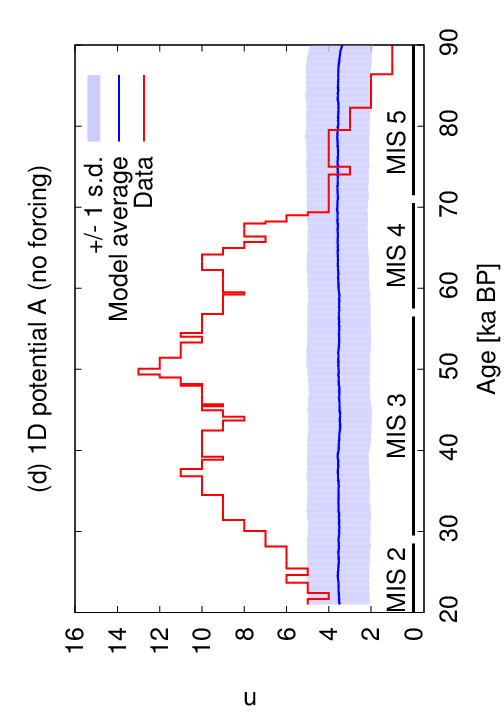}
  \includegraphics[angle=270,width=0.4\linewidth]{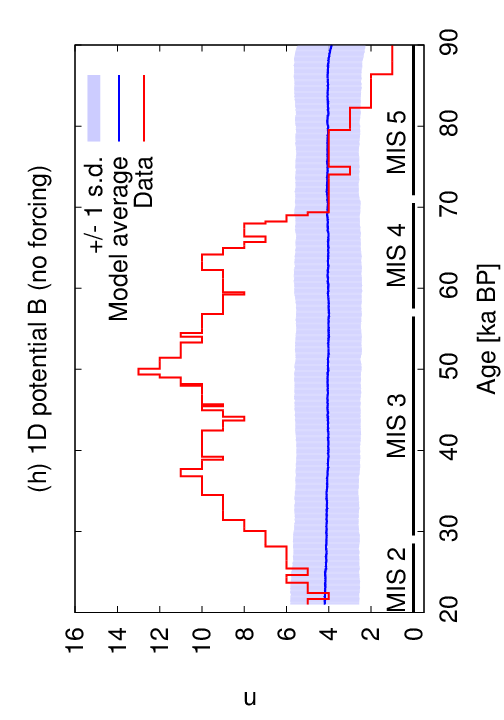}
  \caption{The number of warming transitions $n(t)$ for each 20-ka window [$t-10$~ka, $t+10$~ka]. See text for the definition of a warming transition. The red curve is $n(t)$ of the 20-year average NGRIP Ca$^{2+}$ data, and the blue curve is the ensemble mean $\langle n(t)\rangle$ simulated by each 1D potential model with different noise realizations. The shaded error bar represents $\pm 1$ s.d. Initial conditions are randomly taken at $100$~ka~BP from a normal distribution $p(x)=\mathcal{N} (1,1)$. The ensemble size is $10^3$ for all the models.} \label{fig:nevent_1d}
\end{center}
\end{figure*}

\begin{figure*}
\begin{center}
  \includegraphics[angle=270,width=0.4\linewidth]{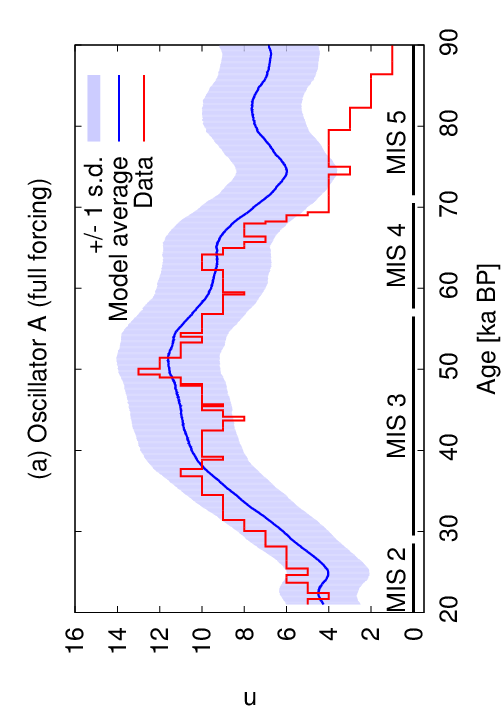}
  \includegraphics[angle=270,width=0.4\linewidth]{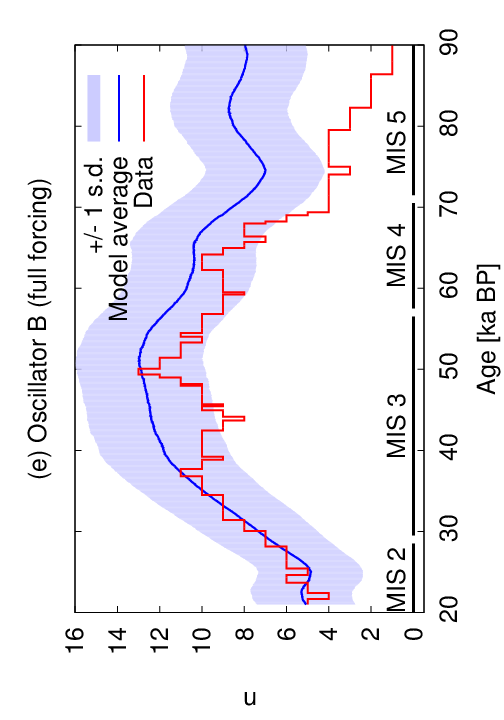}
  \includegraphics[angle=270,width=0.4\linewidth]{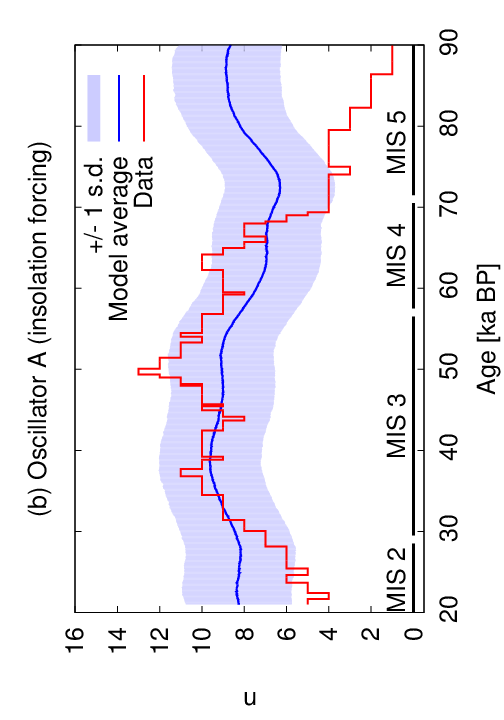}
  \includegraphics[angle=270,width=0.4\linewidth]{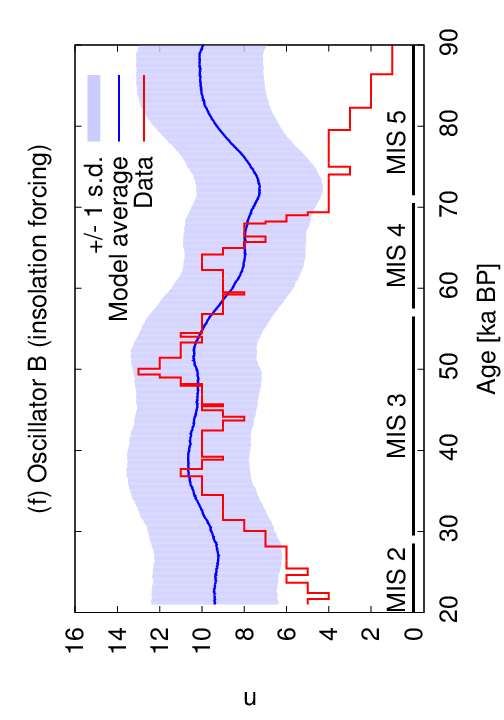}
  \includegraphics[angle=270,width=0.4\linewidth]{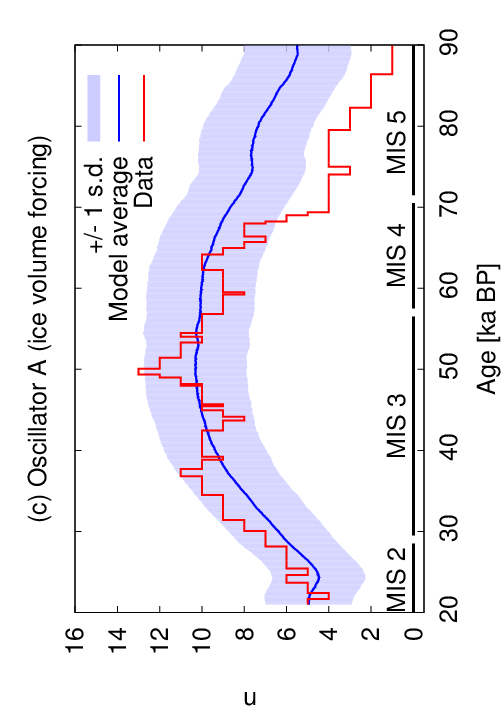}
  \includegraphics[angle=270,width=0.4\linewidth]{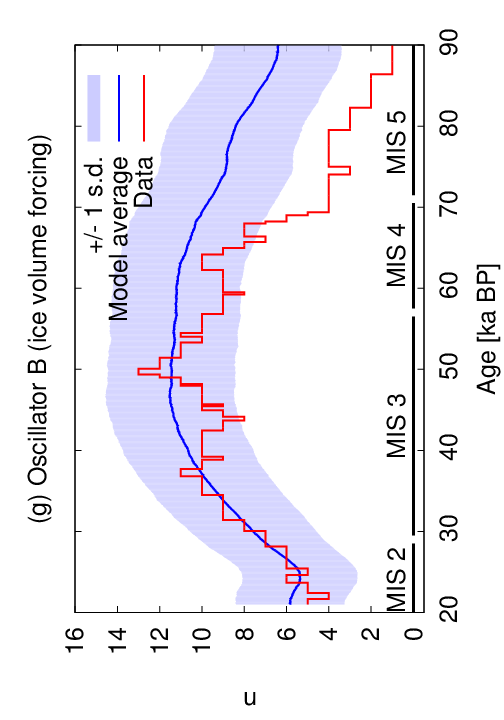}
  \includegraphics[angle=270,width=0.4\linewidth]{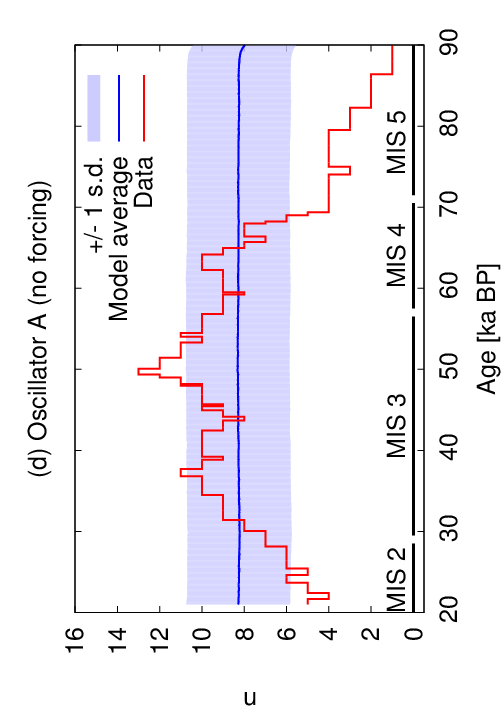}
  \includegraphics[angle=270,width=0.4\linewidth]{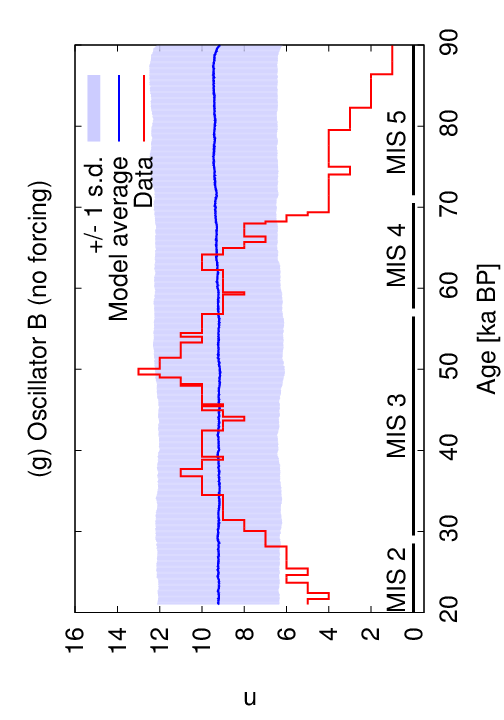}
  \caption{The number of warming transitions $n(t)$ for each 20-ka window [$t-10$~ka, $t+10$~ka]. See text for the definition of a warming transition.  The red curve is $n(t)$ of the 20-year average NGRIP Ca$^{2+}$ data, and the blue curve is the ensemble mean $\langle n(t)\rangle$ simulated by each oscillator model with different noise realizations. The shaded error bar represents $\pm 1$ s.d. Initial conditions are randomly taken at $100$~ka~BP from normal distributions $p(x)= \mathcal{N}(1,1)$ and $p(v)=\mathcal{N} (1,1)$. The ensemble size is $10^3$ for all the models.} \label{fig:nevent}
\end{center}
\end{figure*}

We calculate the average number of warming transitions $\langle n(t)\rangle$ for each 20-ka window [$t-10$~ka, $t+10$~ka] for each model, where $\langle \cdot \rangle$ means the ensemble average for $10^3$ simulations with different noise realizations and initial conditions. The same definition for a warming transition is used as in the Introduction.
Figures~\ref{fig:nevent_1d} and \ref{fig:nevent} show the number of warming transitions $n(t)$ for the NGRIP Ca$^{2+}$ data (red) and the average number of warming transitions $\langle n(t)\rangle$ for each model (blue), where the shaded error bar represents $\pm 1$ s.d.
As mentioned in the Introduction, the number of warming transitions $n(t)$ in the observed record increased from MIS4 to MIS3 and decreased from MIS3 to MIS2. These frequency changes are well reproduced by the oscillator model A with full forcing (Fig.~\ref{fig:nevent}(a)) or with only ice volume forcing (Fig.~\ref{fig:nevent}(c)). The performance of oscillator model B is similar to that of oscillator model A, but the frequency $n(t)$ is larger by about one. These results are robust against small changes of the thresholds ($\pm 0.2$ with respect to $y$).

However, the stochastic oscillator models fail to produce the low values of $n(t)$ in late MIS~5 ($70$--$90$~ka~BP), while this low frequency is somewhat captured by the stochastic 1D potential models with full forcing (Fig.~\ref{fig:nevent_1d}(a)) or with only ice volume forcing (Fig.~\ref{fig:nevent_1d}(c)).

\subsection{BIC and the BIC difference}
{\it 1D potential model vs. oscillator model}. The lowest BIC of the 1D potential model for four different forcing scenarios is $-1375.8$ (Table~\ref{tb:1D2}) and the highest BIC of the oscillator models is $-1743.7$ (Table~\ref{tb:mode1}). Thus, the BIC difference between the worst stochastic oscillator model and the best stochastic 1D potential model is $367.9$. This is very strong evidence in favor of the oscillator models against the 1D potential models. Note that \cite{kwasniok2013analysis} also reports strong evidence in favor of the oscillator model against the 1D potential model for the unforced case $\gamma _i=0$ ($i=0,1,2$) ($\Delta$BIC$=12.9$ for GRIP $\delta ^{18}$O$_\text{ice}$ and $\Delta$BIC$=14.6$ for NGRIP $\delta ^{18}$O$_\text{ice}$).
%
%{\it Oscillator model A vs oscillator model B}. The difference between the lowest BIC of oscillator model A (BIC$=-1748.9$) and that of oscillator model B (BIC$=-1750.4$) is less than $2$. Thus, the evidence in favor of model B against model A is not worth more than a bare mention.

{\it Comparison between different forcings}. As already shown in Tables~\ref{tb:1D1}--\ref{tb:mode2}, the BIC scores suggest that the full forcing or the ice volume forcing are more supported than the insolation forcing or the null forcing. This is consistent with the results obtained by using the sample ACF and the frequency of warming transitions. However, the strength of evidence in favor of a particular forcing depends on the model class as shown in Tables~\ref{tb:bf1}--\ref{tb:bf4}. The evidence in favor of the ice volume forcing against the insolation forcing ($\Delta$BIC) is very strong for the 1D potential models (42.3 for A and 48.4 for B), but it is weak for the oscillator models (1.56 for A and 1.43 for B). On the other hand, the examination of the ACF and the  occurrence frequency of warming transitions suggest qualitatively a more important role for the ice volume forcing than for the insolation forcing. 
%In the next subsection, we compare the models focusing on a specific quantity: the occurrence frequency of DO events.  
%Nevertheless, the model comparison based on the approximate Bayes factor by BIC is consistent with the model comparison based on the frequency of warming transitions in the previous subsection, where the models with ice sheet forcing reproduced the frequency change in MISs~2, 3, and 4, but the model with only insolation forcing cannot reproduce the frequency change well. 
%\begin{table}
%\begin{tabular}{lll}
%\hline\noalign{\smallskip}
%\end{tabular}
%\end{table}
\begin{table*}
\caption{BIC difference $\Delta \text{BIC} _{ij}=\text{BIC} _j-\text{BIC} _i$ as evidence in favor of a model $i$ (row) against a model $j$ (column): the case of the stochastic 1D potential model A. The asterisk means the minus value of corresponding diagonal element.} \label{tb:bf1}
\scalebox{0.7}{
\begin{tabular}{l  c  c  c  c}
\hline\noalign{\smallskip}
              & vs. Full forcing & vs. Insolation only & vs. Ice volume only & vs. No forcing\\
\noalign{\smallskip}\hline\noalign{\smallskip}
Full forcing & $0$ & $48.3$ (very strong) &  $5.96$ (positive) &  $56.2$ (very strong)\\
Insolation only & $\ast$  & $0$ & $\ast$ &  $7.98$ (strong) \\ 
Ice volume only & $\ast$  & $42.3$ (very strong) & $0$ & $50.3$ (very strong)\\
No forcing & $\ast$  & $\ast$ & $\ast$ & $0$ \\
\noalign{\smallskip}\hline
\end{tabular} 
}
\end{table*}
\begin{table*}
\caption{BIC difference $\Delta \text{BIC} _{ij}=\text{BIC} _j-\text{BIC} _i$ as evidence in favor of a model $i$ (row) against a model $j$ (column): the case of the stochastic 1D potential model B. The asterisk means the minus value of corresponding diagonal element.} \label{tb:bf2}
\scalebox{0.7}{
\begin{tabular}{l  c  c  c  c}
\hline\noalign{\smallskip}
              & vs. Full forcing & vs. Insolation only & vs. Ice volume only & vs. No forcing\\
\noalign{\smallskip}\hline\noalign{\smallskip}
Full forcing & $0$ & $55.6$ (very strong) &  $7.18$ (positive) &  $65.7$ (very strong)\\
Insolation only & $\ast$  & $0$ & $\ast$ &  $10.1$ (very strong) \\ 
Ice volume only & $\ast$  & $48.4$ (very strong) & $0$ & $58.5$ (very strong)\\
No forcing & $\ast$  & $\ast$ & $\ast$ & $0$ \\
\noalign{\smallskip}\hline
\end{tabular} 
}
\end{table*}
\begin{table*}
\caption{BIC difference $\Delta \text{BIC} _{ij}=\text{BIC} _j-\text{BIC} _i$ as evidence in favor of a model $i$ (row) against a model $j$ (column): the case of the stochastic oscillator model A. The asterisk means the minus value of corresponding diagonal element.}\label{tb:bf3}
\scalebox{0.7}{
\begin{tabular}{l  c  c  c  c}
\hline\noalign{\smallskip}
              & vs. Full forcing & vs. Insolation only & vs. Ice volume only & vs. No forcing\\
\noalign{\smallskip}\hline\noalign{\smallskip}
Full forcing & $0$ & $3.02$ (positive) &  $1.46$ (weak) &  $5.28$ (positive)\\
Insolation only & $\ast$  & $0$ & $\ast$ & $2.26$ (positive) \\ 
Ice volume only & $\ast$  & $1.56$ (weak) & $0$ & $3.83$ (positive) \\
No forcing & $\ast$  & $\ast$ & $\ast$ & $0$ \\
\noalign{\smallskip}\hline
\end{tabular}
}
\end{table*}
\begin{table*}
\caption{BIC difference $\Delta \text{BIC} _{ij}=\text{BIC} _j-\text{BIC} _i$ as evidence in favor of a model $i$ (row) against a model $j$ (column): the case of the stochastic oscillator model B. The asterisk means the minus value of corresponding diagonal element.}\label{tb:bf4}
\scalebox{0.7}{
\begin{tabular}{l  c  c  c  c}
\hline\noalign{\smallskip}
              & vs. Full forcing & vs. Insolation only & vs. Ice volume only & vs. No forcing\\
\noalign{\smallskip}\hline\noalign{\smallskip}
Full forcing & $0$ & $0.97$ (weak) &  $\ast$ &  $1.32$ (weak)\\
Insolation only & $\ast$  & $0$ & $\ast$ & $0.36$ (weak) \\ 
Ice volume only & $0.47$ (weak)  & $1.43$ (weak) & $0$ & $1.79$ (weak) \\
No forcing & $\ast$  & $\ast$ & $\ast$ & $0$ \\
\noalign{\smallskip}\hline
\end{tabular}
}
\end{table*}

\section{Implication}
The stochastic oscillator model A with full forcing could estimate the occurrence frequency of DO events in the last glacial period (MISs~2--4) though it could not in the early glaciation stage MIS 5. Hence we assume that the model can be extended to past few glacial periods with enough ice sheets, and we predict the frequency of abrupt millennial-scale climate changes in the last four glacial periods under this assumption.

\begin{figure}
\begin{center}
  \includegraphics[angle=0,width=0.9\linewidth]{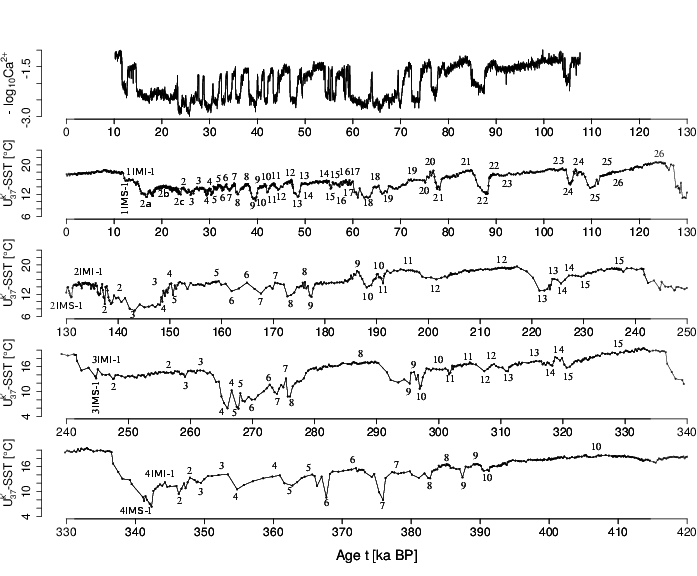}
  \caption{Top panel shows the 20-year average NGRIP Ca$^{2+}$ concentration \citep{rasmussen2014stratigraphic}. The other panels show the U$^{k'}_{37}$--SST in a composite of cores MD01-2444 and MD01-2443 reproduced based on \cite{martrat2007four}. The stage numbers of the Iberian Margin Interstadials (IMI) are shown above the line, and those of the Iberian Margin Stadials (IMS) are shown below the line.
} \label{fig:Martrat2008}
\end{center}
\end{figure}

It is difficult to use Greenland ice core records to infer the frequency of abrupt climate changes before the Eemian interglacial because they are disturbed in chronology due to ice-folding near the bedrock. However, the information may be inferred from Iberian margin SST records derived from the U$^{k'}_{37}$ alkenone index. A U$^{k'}_{37}$-SST record in a composite of cores MD01-2444 and MD01-2443 over the past four glacial cycles is shown in Fig.~\ref{fig:Martrat2008}, which is reproduced based on \cite{martrat2007four}. Similar SST variations are observed also in another Iberian margin core (ODP-997A, not shown here) \citep{martrat2004abrupt}. These U$^{k'}_{37}$-SST records show warmings and coolings corresponding to major DO events in Greenland ice cores.
%, though some of them in the U$^{k'}_{37}$-SST record are too subtle to be identified. 

\cite{martrat2007four} identified cold and warm climate events in the U$^{k'}_{37}$-SST record and labelled them as {\it Iberian margin stadials (IMSs) and Iberian margin interstadials (IMIs)}, respectively. For example, 2IMI-3 denotes the third interstadial within the second glacial cycle (see Fig.~\ref{fig:Martrat2008}).
We use this identification of events to estimate the frequency of abrupt millennial-scale climate changes
in the North Atlantic region though the labels were originally introduced by \cite{martrat2007four} for the purpose of discussion. The timing of each warming transition is set at the time of the largest increase of SST between each IMS and subsequent IMI.

\begin{figure*}
\begin{center}
  \includegraphics[angle=0,width=1\linewidth]{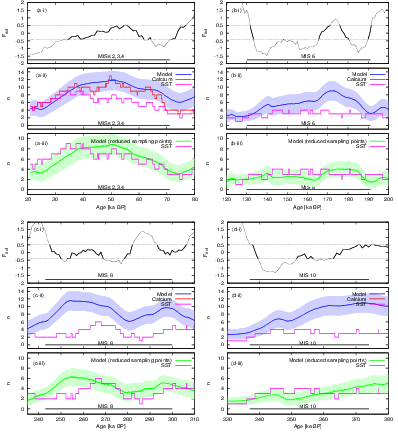}
  %\includegraphics[angle=0,width=1\linewidth]{nevent_ch_resolution_MIS_bind.eps}
%%% convert input.png -trim output.png
  \caption{Simulations of abrupt millennial-scale climate changes in the past four glacial periods (a) MISs 2--4, (b) MIS 6, (c) MIS 8, and (d) MIS 10. (i) Variations of the external forcing $F_\text{ext}(t)=\gamma _1I(t)-\gamma _2V(t)$. Thick parts of the line highlight the range with active stochastic oscillations, $-0.3<F_\text{ext}<0.5$, where the average period between successive warming transitions is less than 2000 years. (ii) The number of warming transitions $n(t)$ for each 20-ka window [$t-10$~ka, $t+10$~ka]. The red curve is $n(t)$ for the 20-year average NGRIP Ca$^{2+}$ data. The magenta line is $n(t)$ for the Iberian margin U$^\text{k'}_{37}$-SST of a composite of cores MD01-2444 and MD01-2443 \citep{martrat2007four}, and the blue curve is the ensemble mean $\langle n(t)\rangle$ simulated by the fully-forced stochastic oscillator model A with different noise realizations. The shaded error bar represents $\pm 1$ s.d. (iii) The green curve is the ensemble mean $\langle n(t)\rangle$ for the simulated time series whose values are sampled at the same time points as the SST recored. Initial conditions are randomly taken at $t=420$~ka~BP from normal distributions $p(x)=\mathcal{N}(1,1)$ and $p(v)=\mathcal{N}(1,1)$.} \label{fig:4cycle}
\end{center}
\end{figure*}

Figures~\ref{fig:4cycle}(a-ii), (b-ii), (c-ii), and (d-ii) show the number of warming transitions $n(t)$ estimated from the Iberian margin SST (magenta) and the average number of warming transitions $\langle n(t)\rangle$ simulated by the stochastic oscillator model (blue). Both are roughly correlated though a phase lag is identified in MIS 8. However, peak levels largely differ between the simulations and the data. We may guess two reasons for the difference: one reason may be that some events do not clearly appear in the SST record. Indeed, some events in Iberian margin SST records seem difficult to be identified without information from Greenland ice cores (for example, 1IMS-14 and 1IMI-13, and 1IMS-23 and 1IMI-22). Another reason may be that some rapid events are missed in the SST record due to its low time resolution especially in the older part.\footnote{For instance, the time intervals between two subsequent data points are $\sim$170~yr on average (with maximum 650~yr) during the past 100~ka~BP, but they are $\sim $480~yr on average (with maximum 1640~yr) during 100-400~ka~BP.} Figures~\ref{fig:4cycle}(a-iii), (b-iii), (c-iii), and (d-iii) (green line) show the average number of warming transitions $\langle n(t)\rangle$ for the simulated time series whose values are sampled at the same time points as the SST record. The number of warming transitions is then similar to that seen in SST data. We therefore suggest that the number of the abrupt climate changes in the past glacial periods was more frequent than that seen in the SST data by \cite{martrat2007four}. 

We now further examine two previous observations. First, \cite{martrat2004abrupt} observed the higher number of abrupt events during MISs 2--4 compared to MIS~6. Consistently, the model simulates a higher average number of warming transitions $\langle n(t)\rangle$ during MISs 2--4 compared to MIS~6 though the difference is smaller in the simulation than in the observation. 
Recall Fig.~\ref{fig:bias}, which shows that stochastic oscillations are frequent when the external forcing $F_\text{ext}(t)=\gamma _1I(t)-\gamma _2V(t)$ is in an intermediate range. Specifically, the average period of oscillations is less than 2000 years for $-0.3<F_\text{ext}<0.5$. As shown in Figs.~\ref{fig:4cycle}(a-i) and (b-i), the external forcing $F_\text{ext}(t)$ is in the intermediate range for short times in MIS~6 but for a long time in MISs 2--4. This explains why the number of warming transitions $n(t)$ during MISs 2--4 is higher than during MIS~6.

Secondly, \cite{martrat2007four} consider that the millennial-scale climate changes become abundant as the Pleistocene progresses to the present. However, if the model can be extended to the older glacial periods, millennial-scale climate changes in MIS~8 and MIS~10 are expected to be as frequent as in the last glacial period, as shown in Figs.~\ref{fig:4cycle}(c-ii) and (d-ii). 

\section{Conclusion}
The influence of external forcings on DO events was investigated with statistical modeling based on simple stochastic dynamical systems: the 1D potential model and the oscillator model forced by the northern hemisphere summer insolation change and the global ice volume change. We estimated model parameters by maximizing the likelihood with the NGRIP Ca$^{2+}$ record. The stochastic oscillator model at least with the ice volume forcing reproduces well the sample autocorrelation function of the record and the frequency changes of warming transitions in the last glacial period across MISs~2--4. The model performance is improved with the additional insolation forcing. The BIC scores also suggest that the ice volume forcing is relatively more important than the insolation forcing, though the strength of evidence ($\Delta$BIC) depends on the model assumption on the system and noise. 

It is worth mentioning that not only the influence of the insolation forcing but also the influence of the ice volume forcing is detected in grain-size records from the Chinese Loess Plateau (a proxy for the East Asian winter monsoon intensity) by spectral analyses \citep{ding1995ice,li2015multiscale}. This is consistent with our result, given the proximity of the sources of Chinese loess archives and the sources of terrestrial dusts found in Greenland.

Finally, using the fully-forced oscillator model A calibrated in the last glacial cycle, we simulated the average number of warming transitions $\langle n(t)\rangle$ for each 20-ka moving window over the past four glacial periods, and compared the result with an Iberian margin SST data \citep{martrat2007four}. The simulation result supports the previous observation by \cite{martrat2004abrupt} that abrupt millennial-scale climate changes in the penultimate glacial (MIS~6) is less frequent than in the last glacial (MISs~2--4). On the other hand, it suggests that the number of abrupt millennial-scale climate changes in older glacial periods (MISs 6, 8, and 10) might be larger than inferred from the SST data. If the model can be extended to the older glacial periods, the millennial-scale climate changes in MIS~8 and MIS~10 are expected to be as frequent as in the last glacial period.    

The LR04 record, used here as a proxy for the global ice volume, contains information about astronomical insolation forcing. Hence, the ice volume forcing implicitly includes a component of the insolation forcing. It is therefore not so surprising to find only weak evidence for the need of additional insolation forcing. We should be cautious not to interpret our result as evidence that the global ice volume is the only physical factor controlling the frequency changes of DO events. 

The calibrated models did not reproduce well the low occurrence frequency of DO events in the last glaciation period (MIS~5) (Fig.~\ref{fig:nevent}). A use of another ice volume estimate \citep{bintanja2005modelled} modifies the discrepancy slightly but not substantially (data not shown). To overcome this, we would need to examine the model assumptions introduced for simplicity.  The orography of ice sheets might be more effective than the global ice volume \citep{zhang2014abrupt}. Multiplicative effects of the insolation and the ice volume might be important, given the ice-albedo feedback. A state-dependent noise may have to be considered \citep{ditlevsen1999observation,timmermann2000noise}. Indeed, fluctuations in $\delta ^{18}$O$_{\text{ice}}$ have larger variances in stadial than in interstadial \citep{ditlevsen2002fast}. Non-Gaussian noise and/or temporally-correlated noise might also be suitable to represent external disturbances (such as massive iceberg discharges) \citep{ditlevsen1999observation}. 

\begin{acknowledgements}
We thank P.~Ditlevsen, H.~Goosse, M.~Van~Ginderachter, D.~Kondrashov, and E.~W.~Wolff for helpful comments and suggestions. This work is supported by the Belgian Federal Science Policy Office under contract BR/12/A2/STOCHCLIM. MC is research scientist with the Belgian National Fund of Scientific Research.
\end{acknowledgements}

\bibliographystyle{spbasic}      % basic style, author-year citations
%\bibliography{../mybib.bib}   % name your BibTeX data base
\bibliography{mybib}   % name your BibTeX data base

%\bibliographystyle{plainnat}

%\begin{thebibliography}{200}
% etc
%\end{thebibliography}
%\end{linenumbers}
\end{document}